\pdfoutput=1
\documentclass[twocolumn,a4paper]{article} 
\usepackage{graphicx}
\usepackage{color}
\usepackage{mathrsfs}
\usepackage{xifthen}
\usepackage{amssymb}
\usepackage{amsmath}
\usepackage{textgreek}
\usepackage{xstring}
\usepackage[normalem]{ulem}
\usepackage{cite} 
\usepackage[font=small]{caption}
\usepackage{cite}
\newcommand{\Ca}{$^{40}$Ca$^+$}
 
\newcommand{\ket}[1]{\left| #1 \right\rangle}
\newcommand{\bra}[1]{\left\langle #1 \right|}
\newcommand{\state}[3]{\ensuremath{
	\ifthenelse{\isempty{#3}{}}
		{\ket{#1_{#2}}}
		{\ket{#1_{#2},#3}}}%
}
\DeclareMathOperator{\erf}{erf}

\usepackage[pdftex,colorlinks,bookmarks=false]{hyperref}
\begin{document}
\title{A Quantum Repeater Node with Trapped Ions: A Realistic Case Example}
\author{A. D. Pfister$^{1\ast}$, M. Salz$^{1}$, M. Hettrich$^{1}$, U. G. Poschinger$^{1}$, F. Schmidt-Kaler$^{1}$
}
%
%

%
\date{16.04.2016}
%
\maketitle
\begin{abstract}
We evaluate the feasibility of the implementation of two quantum repeater protocols with an existing experimental platform based on a \Ca-ion in a segmented micro trap, and a third one that requires small changes to the platform. 
A fiber cavity serves as an ion-light interface. 
Its small mode volume allows for a large coupling strength of $g_c = 2 \pi \times 20$~MHz despite comparatively large losses $\kappa = 2\pi \times 18.3$~MHz.
With a fiber diameter of $125$~\textmu m, the cavity is integrated into the microstructured ion trap, which in turn is used to transport single ions in and out of the interaction zone in the fiber cavity. 
We evaluate the entanglement generation rate for a given fidelity using parameters from the experimental setup. 
The  DLCZ protocol~\cite{Duan2001} and the hybrid protocol~\cite{Loock2006} outperform the EPR protocol~\cite{1367-2630-15-8-085004}.
We calculate rates of more than than 100~s$^{-1}$ for non-local Bell state fidelities larger than 0.95 with the existing platform. 
We identify parameters which mainly limit the attainable rates, and conclude that entanglement generation rates of 750~s$^{-1}$ at fidelities of 0.95 are within reach with current technology.

\end{abstract}

\section{Introduction}
\label{sec:intro}
One of the most prominent applications in quantum technologies is quantum key distribution (QKD).
The fundamental no-cloning theorem of quantum states~\cite{Wooters1982} enables secure communication protocols~\cite{Bennett1984,Ekert1991}.
For distances less than 80~km QKD is commercially available~\cite{IDQuantique,magiQ} using standard telecom fiber networks, and QKD networks have been set up in multiple locations, e.g. near Tokyo, Vienna and Boston. 
However, the transmission losses of 0.2~dB/km in fibers lead to restrictions at higher distances, as the secure key rate drops exponentially with the fiber length. 
Two potential solutions are currently being discussed: 
Free-space optical links have been established and tested between two of the Canary Islands~\cite{Ma2012a}, and between a ground station and a satellite~\cite{Vallone2015}. 
On the other hand, the abundance and maturity of fiber optical networks appears to be appealing for use in QKD. 

\begin{figure}
	\centering
		\includegraphics[width=.47\textwidth]{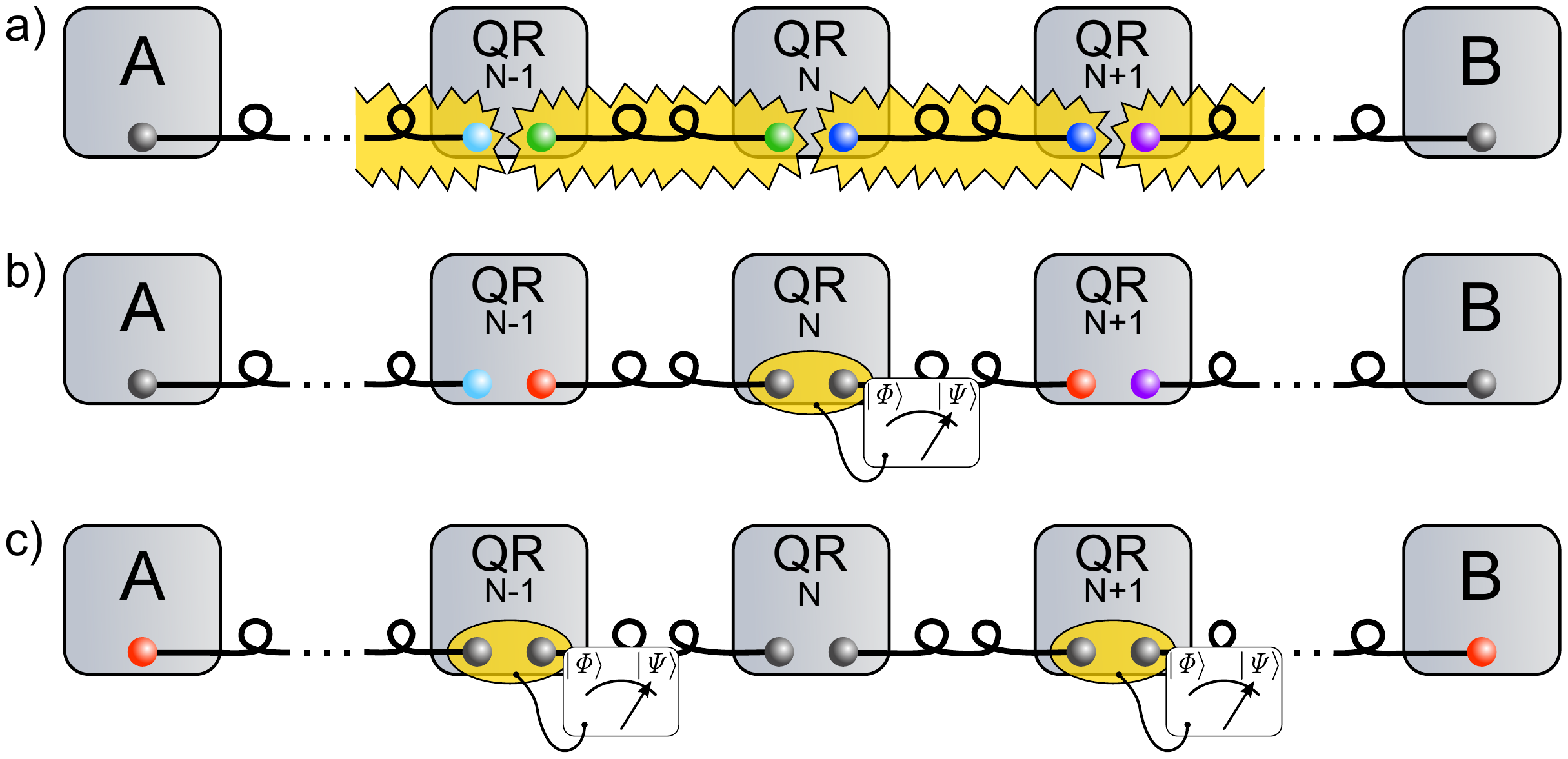}
	\caption{Scheme of long distance QKD using quantum repeaters (QR): 
a) Entanglement (indicated by colors) is generated between stationary qubits of adjacent repeater nodes via optical channels (black lines). 
b) A local Bell state measurement allows for an entanglement swapping procedure.
c) Consecutive entanglement swapping operations ultimately leave the system with Alice (A) and Bob (B) having one entangled Bell pair.
Finally, the Eckert protocol~\cite{Ekert1991} allows for a secret QKD between Alice and Bob.
}
	\label{fig:QrepPrinciple}       
\end{figure}

The proposal by Briegel, D\"ur, Cirac and Zoller~\cite{Briegel1998} overcomes the distance limitation by converting the exponential drop in key rates to a polynomial one by using a network of \emph{quantum repeaters} (QR).
Here, entanglement is generated at QRs separated by a shorter distance where photonic channels are still efficient (Fig. \ref{fig:QrepPrinciple}). 
By a sequence of entanglement swapping operations, entanglement is generated between the distant endpoints, commonly referred to as \textit{Alice} and \textit{Bob}. 
These entangled qubits allow Alice and Bob to perform QKD \cite{Ekert1991}.

The building blocks of a QR following Briegel et al. (BDCZ-QR) are (i) an efficient interface between a flying (photonic) qubit and a long-lived stationary quantum memory, (ii) quantum logic operations on the memory, and (iii) error correction protocols \cite{Bennett1996,Calderbank1996,Reichle2006,Terhal2015}.
Practical considerations also suggest (iv) a wavelength-transformer such that the transmitted photons are near telecom wavelengths around 1.5 \textmu m, where fiber losses are minimal.
An alternative proposal \cite{Munro2012} relaxes the requirement for long-lived quantum memories, but at the price of a substantially increased number of required qubits.

A wide variety of possible implementations is currently being investigated~\cite{Pirandola2015}, based either on atomic systems and quantum optical techniques~\cite{Kuzmich2003,Duan2010,Ritter2012,Hucul2014} or on solid state quantum devices~\cite{Press2008,Gschrey2015}.
However, up to date there has been no demonstration of a fully functional QR.

Here, we focus on a trapped ion approach: advantages are the high fidelity gate operations and state readout, and a long coherence time. 
Many of the requirements for (ii) are met and modern ion trap technology allows for scaling up to modest numbers of qubits required for a QR.
Light-atom interfaces (i) in the quantum regime have been demonstrated by placing ions into high-finesse optical cavities~\cite{Mundt2002,Steiner2014,Casabone2015}. 
Furthermore, single photon conversion to telecom wavelengths (iv) has recently been demonstrated~\cite{Ikuta_2011,Zaske2012}.
However, the scalable combination of (i - iv) still appears to be very demanding technologically.

The scope of this work is a detailed investigation of the practical joint implementation of an ion-light interface (i) and quantum logic operations (ii) basic building blocks required for a BDCZ-QR.
We do not address wavelength conversion (iv), fiber losses, classical verification and reconciliation protocols \cite{Brassard1994}, or privacy amplification \cite{Bennett1988,Deutsch1996}.
While (iii) error correction is closely related to the subject of this manuscript, we only focus on the elementary link.
We make a detailed comparison between three different entanglement distribution protocols and to identify critical experimental parameters limiting the performance.

After outlining the protocols in general (Sec. \ref{sec:prot}), we describe the trapped ion experiment and its key parameters (Sec. \ref{sec:setup}), and describe how to implement each of the protocols on the trapped-ion platform (Sec. \ref{sec:assess}). 
Finally, we specify the attainable entanglement generation rates for each protocol. 
Based on this, we identify the protocols which are most suitable on our specific experimental platform.

%

\section{Quantum repeater protocol options}
\label{sec:prot}

In this work, we will investigate the following entanglement distribution protocols:
\begin{itemize}
\item a scheme using \emph{distributed Einstein-Podolski-Rosen-(EPR) states} from a source of entangled photons~\cite{1367-2630-15-8-085004,Lloyd2001}
\item two variations of the \emph{Duan-Lukin-Cirac-Zoller-\\(DLCZ) protocol}~\cite{Duan2001,Cabrillo1999,Simon2003}
\item a protocol using a combination of discrete and continuous variables dubbed \emph{hybrid protocol} (HP)~\cite{Loock2006}
\end{itemize}

The key parameters for the assessment of the protocols are the fidelities $F$ and rates of successful entanglement per second $r_e = P / t_d$, where $P$ is the success probability, and $1/t_d$ the repetition rate. 
The fidelities and rates of entanglement for all three protocols are inferred using experimentally determined or estimated parameters of the apparatus.

All the protocols have in common that they use photonic flying qubits, although the hybrid protocol is unique in that entanglement is distributed using a continuous variable encoded in a coherent light pulse instead of polarization entangled photons.
Furthermore, all protocols feature heralding of entanglement creation.

\subsection{Distributed EPR-states protocol}
\label{EPR}

\begin{figure}
	\centering
		\includegraphics[width=0.48\textwidth]{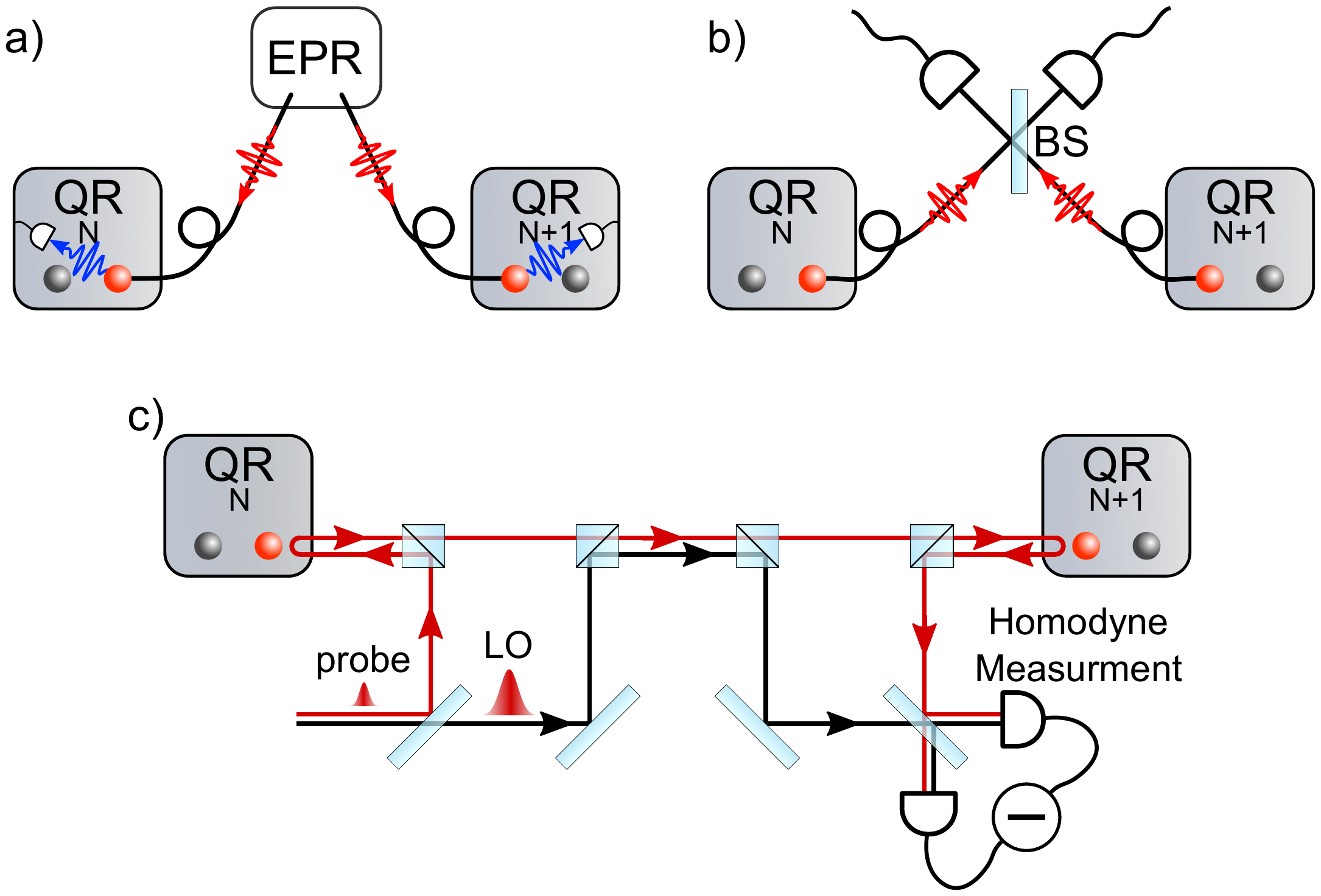}
	\caption{Basic principle for distant entanglement generation for the three protocols investigated. Stationary qubits with shared entanglement are indicated as red balls.
\\ \textbf{a)}
Distributed EPR-states protocol.
An entangled photon pair (red waves) is generated by an EPR source and sent to two adjacent repeater nodes. 
The quantum state of the photons is then mapped on a stationary qubit at each node, leaving one qubit at each node entangled with each other after detection of the herald photon (blue) 
\\ \textbf{b)}
DLCZ protocol.
A local operation at each repeater node probabilistically generates a photon in either of the nodes.
Interfering the possible photon paths on a beam splitter (BS) before detection entangles the qubits.
\\ \textbf{c)}
Continuous variables hybrid protocol, adapted from~\cite{Loock2006}.
A coherent light pulse is split into a weak probe pulse (qubus) and a  strong local oscillator (LO). 
The qubus interacts dispersively with the stationary qubit, resulting in a state dependent phase shift of the qubus. 
After transferring the qubus via the optical channel, the same operation is performed at the second repeater station.
Measurement of the qubus phase leaves the system in an non-maximally entangled Bell state, for a correct detection pattern.
}
	\label{fig:theo}       
\end{figure}

One of the most prominent schemes proposed for quantum communication through multiple repeater stations is based on the idea of distributing the constituents of an entangled photon pair, e.g. polarization entangled photons from a spontaneous down conversion source. These are transmitted via fibers to neighboring QR nodes (N) and (N+1), c.f. Fig. \ref{fig:theo}a)\,.
There, the photon state is mapped onto stationary qubits, giving rise to inter-node entanglement~\cite{Lloyd2001}. 
Building blocks of this scheme have been realized e.g., in~\cite{Kurz2014,Schug2014}.

For this protocol, the stationary qubits are initially prepared in a superposition state, $\ket{\Psi_i}_q=\ket{0_i} + \ket{1_i}$. 
Throughout the paper, we omit the normalization of wavefunctions, unless the normalization factor is of specific interest.

A polarization-entangled photon pair in the state $\ket{\Psi}_p$ interacting with the stationary qubit couples both states $\ket{0_i}$ and $\ket{1_i}$, depending on its polarization, to levels of a short lived, excited state, $\ket{0_e}$ and $\ket{1_e}$: 
\begin{align}
\label{eq:epr:abs}
	\ket{\Psi_i}_q^{\otimes 2} \otimes \ket{\Psi}_p =&  (\ket{0_i} + \ket{1_i})^{\otimes 2} \big(  \ket{\sigma^+,\sigma^-} +\ket{\sigma^-,\sigma^+} \big) \nonumber \\
	\xrightarrow{\text{\quad abs.\quad}}& \ket{0_e,1_e} + \ket{1_e,0_e} \nonumber \\
	+& \big(\ket{0_e,0_i} + \ket{0_i,0_e} \big) \ket{\sigma^-} \nonumber \\
	+& \big(\ket{1_e,1_i} + \ket{1_i,1_e} \big) \ket{\sigma^+} \\
	+&\ket{0_i,1_i}\ket{\sigma^-,\sigma^+} + \ket{1_i,0_i}\ket{\sigma^-,\sigma^+} \nonumber
\end{align}
The excited states decay into a long-lived state, $\ket{0_f}$ and $\ket{1_f}$, leaving an entangled final state upon two-photon emission:
\begin{align}
\label{eq:epr:em}
	\ket{\Psi_f}_q=\big(\ket{0_f,1_f} + \ket{1_f,0_f}\big).
\end{align}
This decay gives access to herald photons via the spontaneous emission.

Both the initial states and the herald detection basis have to be chosen such that the decay via the distinct channels $\ket{0_e}\rightarrow\ket{0_f}$ and $\ket{1_e}\rightarrow\ket{1_f}$ is guaranteed. At the same time, the availability of which-path-information has to be prevented for preserving entanglement.

Due to a low absorption efficiency, in most cases either only one or zero photons of an EPR pair interacts at a node.
We are interested in the probability of a single-photon-interaction taking place at either node within time $t$, after an initialized stationary qubit is exposed to the EPR source, which is given by
\begin{equation}
\label{eq:eprP1}
	P_1(t) = 1-e^{-r_1 \cdot t}.
\end{equation}
The rate of single herald photon emission events
\begin{equation}
\label{eq:eprr1}
	r_1 = r_\text{EPR} \cdot \eta
\end{equation}
depends on the brightness of the EPR source, $r_\text{EPR}$, and the probability $\eta$ of a photon from the EPR source to be injected into a cavity at a node and to interact with the stationary qubit.

The probability density for both photons of one EPR pair to interact each with its stationary qubit at time $t$, while \textit{no} single-photon-interaction at neither node has happened before, is
\begin{equation}
\label{eq:epr2dens}
	p_2(t) = r_2 \, e^{-r_2 \cdot t}\, (1 - P_1(t))^2 = r_2\, e^{-(r_2 + 2 \,r_1) \cdot t}.
\end{equation}
The rate of \emph{both} photons from one EPR pair interacting with their respective stationary qubit is given by
\begin{equation}
\label{eq:eprr2}
	r_2 = r_\text{EPR} \eta^2.
\end{equation}

In Eq. \ref{eq:epr2dens} we explicitly exclude a single photon event taking place during time $t$, since such an event would change the state of a stationary qubit and thwart any two-photon event.

Integrating Eq. \ref{eq:epr2dens} over $t$, we find the total probability of a successful two-photon mapping within time $t$ after initialization to be
\begin{equation}
\label{eq:eprP2}
	P_2(t) = \int_0^t{p_2(\tau) d\tau} = \frac{r_2}{r_2+2\, r_1}\big(1-e^{-(r_2+2 r_1)t}\big).
\end{equation}

The emitted heralds are detected at a probability $P_\text{det}$, leading to a success probability for one experimental run of
\begin{equation}
\label{eq:eprheraldprob}
	P_e(t) = P_\text{det}^2 P_2(t).
\end{equation}
Thus, the entanglement generation rate of stationary qubits at neighboring QR nodes is given by
\begin{equation}
\label{eq:eprRate}
	r_e = \frac{P_e(\tau_W)}{\tau_W+\tau_\text{prep}},
\end{equation}
where $\tau_W\approx \frac{1}{2r_1}$ is the detection window for coincident herald photons. 
After expiry of $\tau_W$ without coincident herald detection, the stationary qubits at both nodes are re-initialized, which takes the time $\tau_\text{prep}$. 
Thus, for fixed $r_1$ and $\tau_\text{prep}$, $\tau_W$ can be chosen to provide an optimum entanglement rate.
%

\subsection{DLCZ-protocol}
\label{DLCZ}
%

This repeater protocol entangles stationary qubits at different repeater nodes by probabilistically creating a stationary qubit-photon pair in either of the nodes (N) and (N+1), see Fig. \ref{fig:theo}b).
A detection registering the arrival of one photon, but unable to distinguish the source node, projects the stationary qubits at the two nodes into a Bell state. 

The stationary qubits are initialized in a state $\ket{q_0}$.
A laser beam then drives a cavity-induced stimulated Raman transition~\cite{Stute2012} via the excited state $\ket{q_e}$ to a stable, final state $\ket{q_1}$.
The quantum state of the stationary qubit and the cavity mode at one node evolves as
\begin{equation}
\label{eq:dlczevo}
	\ket{q_0}\ket{0_c}\xrightarrow{\text{Raman}}\ket{q_0}\ket{0_c}+\sqrt{p_1}\ket{q_1} \ket{1_c}.
\end{equation}
Here, $\ket{n_c}$ is the $n$-photon Fock state of the cavity mode.

The entanglement between neighboring nodes is a\-c\-h\-iev\-ed  by a \emph{single-photon detection scheme}~\cite{Cabrillo1999}.
Driving the Raman transition such that the transition probability fulfills $p_1 \ll 1$, the state evolution in both nodes is
\begin{equation}
	(\ket{q_0}\ket{0_c})^{\otimes 2} \rightarrow \ket{q_0, q_1}\ket{0_c,1_c} + \ket{q_1,q_0}\ket{1_c,0_c},
\end{equation}
where we have omitted the parts of the final wave function where either no or two photons is emitted.
The emitted photons are transmitted to a detection setup, where a 50/50 beamsplitter in front of two detectors erases which-path information, see Fig. \ref{fig:theo} b.
This scheme results in the two-qubit wave function
\begin{equation}
	 \big( \ket{q_1,q_0} + \ket{q_0,q_1} \big) \ket{A}+ \big( \ket{q_1,q_0} - \ket{q_0,q_1} \big) \ket{B},
\end{equation}
where either detector $A$ or $B$ registers a photon.

The success probability of this entanglement creation, with single-photon emission probability $p_1$ and probability to detect the emitted photon $P_\text{det}$, is given by~\cite{Zippilli2008}
\begin{align}
\label{eq:dlcz1suc}
	P_{1}=& 2\, P_\text{det}\, p_1\, ( 1- P_\text{det}\, p_1).
\end{align}
The two-photon emission process omitted in Eq. \ref{eq:dlczevo} leads to an infidelity of entanglement generation.
Therefore, $p_1$ has to be chosen sufficiencly small in order to reach a given threshold fidelity $F_\text{thr}$, 
\begin{equation}
\label{eq:dlcz:1suc}
	p_1 \leq \frac{1-F_\text{thr}}{1-P_\text{det} F_\text{thr}}.
\end{equation}

The entanglement generation rate $r_e$ with success rate $P_1$ can be found by dividing $P_1$ by the required time per experimental run $\tau_\text{run}$:

\begin{equation}
\label{eq:dlczrate}
	r_e = \frac{P_1}{\tau_\text{run}}.
\end{equation}

\subsection{Continuous variables hybrid protocol}
\label{sub:Hyb}

The two previous protocols operate on discrete variables, both for the flying and the stationary qubits. In contrast, the hybrid protocol~\cite{Loock2006,Ladd2006,Loock2008} employs continuous variables for encoding photonic quantum information, while retaining the discrete stationary qubit, see Fig. \ref{fig:theo}c).

The continuous variable is embodied by a coherent light pulse, termed \textit{qubus}.
The stationary qubits are initially prepared in a superposition state $\ket{\Psi_i}=\ket{0}+\ket{1}$, and interact with the qubus, which is injected into a cavity, to enhance the interaction.
The cavity field off-resonantly drives the transition $\ket{1}\leftrightarrow\ket{e}$ to an auxiliary excited state. 
The detuning $\Delta$ from this transition is much larger than the vacuum Rabi splitting, $\Delta\gg 2g$, such that the interaction is \textit{dispersive}.
The state of the stationary qubit is imprinted into the phase of the qubus state $\ket{\alpha}$.
The Hamiltonian pertaining to this regime is given by
\begin{equation}
	\hat{H}_\text{int} = \hbar \frac{g^2}{\Delta} \hat{\sigma}_z \hat{a}^\dagger \hat{a}.
\end{equation}
The operator $\hat{a}^\dagger$ ($\hat{a}$) is the creation (annihilation) operator of the field mode, and $\hat{\sigma}_z = \ket{0}\bra{0}-\ket{1}\bra{1}$ is the Pauli z-operator.
This Hamiltonian describes an energy shift  dependent on the state of the stationary qubit.
The evolution operator of this Hamiltonian is
\begin{equation}
	\hat{U}_\text{int} = \exp{ \Bigg(-i \frac{\theta}{2} \hat{\sigma}_z \hat{a}^\dagger \hat{a}\,  \Bigg) } ,
\end{equation}
with a phase shift
\begin{equation}
	\theta = \frac{2 \, g^2}{\Delta}\, \tau_\kappa,
\end{equation}
where $\tau_\kappa$ is the interaction time in the cavity.
For a coherent state in the cavity, this leads to the following evolution of a superposition state of the stationary qubit:
\begin{equation}
	\hat{U}_\text{int}\big( \ket{0} + \ket{1} \big)\ket{\alpha} = \ket{0}\ket{\alpha  e^{-i\theta/2}}+\ket{1}\ket{\alpha e^{i\theta/2}}.
\end{equation}

Neglecting losses, the interaction of the qubus with two stationary qubits in neighboring QR nodes leads to the state~\cite{Loock2006,Loock2008}
\begin{align}
	\ket{\Psi}=\ket{\psi^+}\ket{\alpha}& + \ket{00}\ket{\alpha e^{-i\theta}} + \ket{11}\ket{\alpha e^{i\theta}},
\end{align}
A measurement determining the phase of the qubus pro\-jects the state into either one of the three components. 
The Bell state $\ket{\psi^+}$ is generated in the QR nodes if no phase shift is detected.

We thus require that the \textit{distinguishability} 
\begin{equation}
\label{eq:hyb:d}
	d=\alpha \sin{\theta}
\end{equation} of the phase shifted states is sufficiently large to separate the coherent states in phase space~\cite{Loock2006}.

While a large amplitude of $\ket{\alpha}$ will guarantee a high distinguishability, it also leads to decoherence when losses in the transmission of the qubus are taken into account..
One thus faces a trade off between fidelity and efficiency.

Taking losses into account, we introduce a total transmission $\eta$ from the cavity in QR (N) to the cavity in QR (N+1), so that on average $(1-\eta)|\alpha|^2$ photons will be lost to the environment while the qubus propagates between neighboring nodes.
Following~\cite{Loock2008}, we define the coherence parameter
\begin{equation}
  	\label{eq:hybF}
	\mu^2 = \frac{1}{2} \big( 1+e^{-(1-\eta) \alpha^2 (1-\cos{\theta})} \big).
\end{equation}
The initial pure state evolves to a mixed state, after a local operation on each qubit, with a density matrix
\begin{equation}
\label{eq:hyb:state}	
	\ket{\Psi}\bra{\Psi} = \mu^2 \ket{\Psi^+}\bra{\Psi^+}+(1-\mu^2)\ket{\Psi^-}\bra{\Psi^-},
\end{equation}
where
\begin{align}
\label{eq:hyb:phip}
	\ket{\Psi^{\pm}} =& \frac{1}{\sqrt{2}}\ket{\sqrt{\eta}\alpha}\ket{\psi^{\pm}} \pm \frac{1}{2}e^{-i\eta\xi}\ket{\sqrt{\eta}\alpha e^{i\theta}}\ket{11} \\
	& + \frac{1}{2} e^{i\eta\xi}\ket{\sqrt{\eta}\alpha e^{-i\theta}}\ket{00}. \nonumber 
\end{align}

The relative phase $\xi$ is of no further significance for our discussion.
Projecting the wave function of the qubus to the non-phase-shifted part at detection selects the maximally entangled Bell states $\ket{\psi^\pm} =  \ket{10} \pm\ket{01} $.
A high distinguishability implies reliable identification of the Bell states, but mixes the two pure states $\ket{\Psi^\pm}$, whereas for low distinguishability, a pure state $\ket{\Psi^+}$ is dominant, at the price of reduced success when identifying the Bell state $\ket{\psi^+}$.

In order to collapse the wave function to the required part, it is necessary to identify the phase of a coherent state $\ket{\alpha}$. 
One possibility is \emph{p-homodyne detection}, which requires a setup as depicted in Fig. \ref{fig:theo} c.
In this setup, the reference signal for homodyning is created by splitting a coherent pulse into the weak qubus signal and a local oscillator (LO) phase reference pulse, which does not interact with the cavities.
The p-homodyne measurement amounts to a projection of the qubus state onto the p-quadrature of phase space. 
Following~\cite{Ladd2006}, for an acceptance window $-p_c<p<p_c$, we can assign a probability $P_S$ of a 'non-phase-shifted' detection event and a fidelity $F$ of the resulting state,

\begin{align}
\label{eq:Phomo}
	P_S =& \frac{1}{4} \Big(2 \erf{\big(\sqrt{2}p_c\big)}+\erf{\big(\sqrt{2}(p_c+\eta d)\big)} \\ 
	& +\erf{\big(\sqrt{2}(p_c - \eta d)\big)}\Big) \nonumber \\
\label{eq:Fhomo}	
	F =& \bra{\psi^+}\rho\ket{\psi^+} \nonumber  \\
	=&\frac{1}{4 P_S}\big(1 + e^{-d^2 (1-\eta^2)/2} \big)\erf{\big(\sqrt{2}p_c\big)}.
\end{align}

In each of the presented protocols, a Bell state is generated first between nodes QR (N) and QR (N+1), and then between QR (N) and QR (N$-$1) in the network (Fig. \ref{fig:QrepPrinciple} a). 
Entanglement swapping~\cite{Bennett1993 ,Riebe2008} then creates Bell states in nodes QR (N$-$1) and QR (N+1), which have twice the distance (Fig. \ref{fig:QrepPrinciple} b).
The repeated application of entanglement swapping finally leads to entanglement between the end nodes Alice and Bob.

\section{Experimental platform}
\label{sec:setup}

In this section, we describe the relevant components of our ion trap / cavity setup. 
We explicitly give quantitative parameters of this experimental platform, which are used in Sec. \ref{sec:assess} to assess the performance of the different QR  protocols.
All the following components  have been demonstrated to work as described in our labs. 
A single apparatus with all required parts integrated is not yet operative, the reported values thus pertain to similar trap apparatuses operated in our laboratories \cite{Schulz2008, Poschinger2009}.

\subsection{Segmented microtrap}
\label{sub:trap}

\begin{figure}
		\includegraphics[width=.48\textwidth]{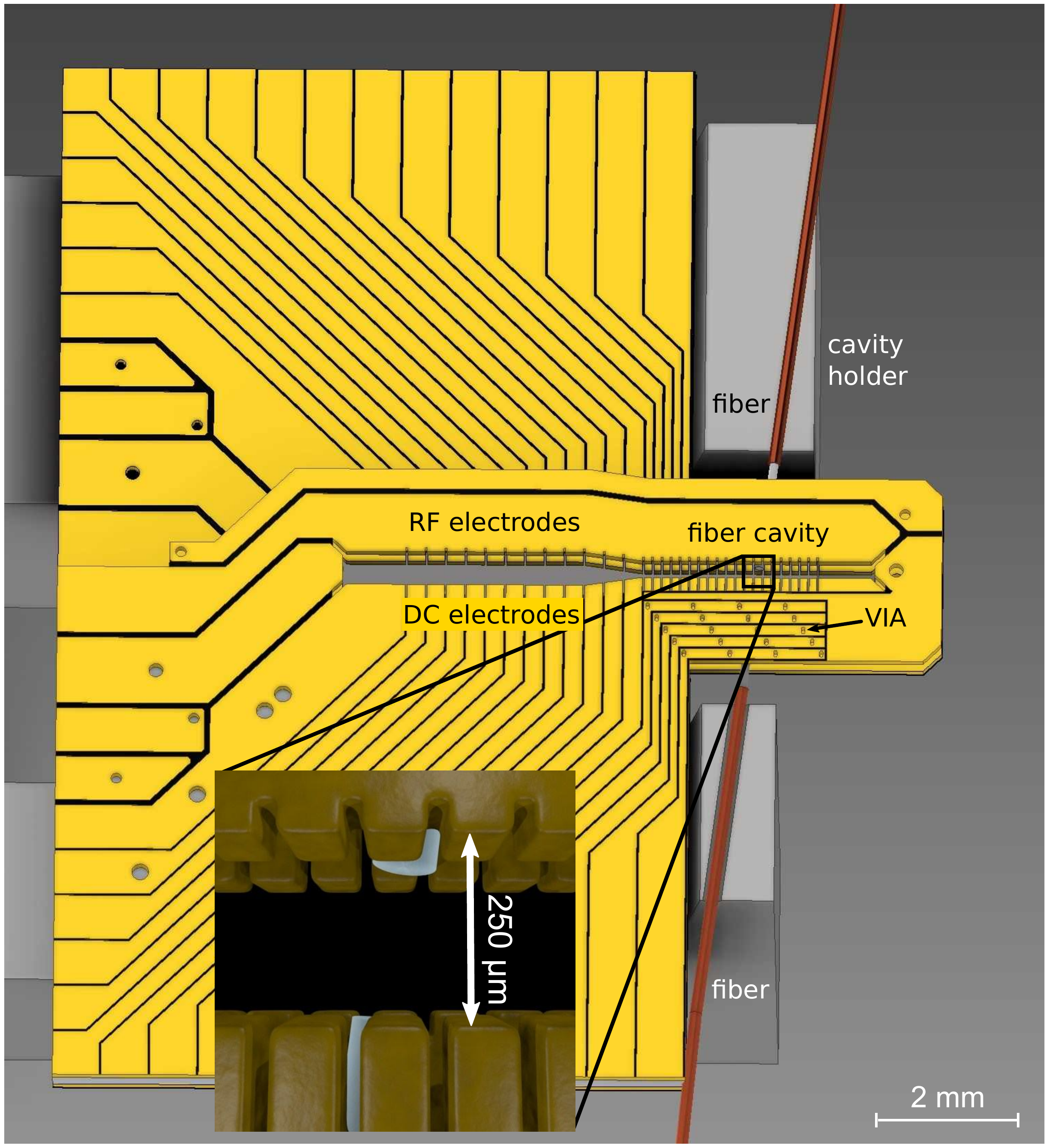}
	\caption{Sketch of the microstructured trap bearing the fiber cavity (see inset). 
	The trap design provides access for the fiber cavity in the small processing region.
	The design of the trap assembly allows the cavity to be flexibly placed with respect to the trap axis.}
	\label{fig:trap}       
\end{figure}

Our setup consists of a variation of the segmented, microstructured Paul trap described in~\cite{Schulz2006,Schulz2008}, adapted to accommodate a fiber based cavity similar to~\cite{Hunger2010} as depicted in Fig. \ref{fig:trap}.
The trap segments are laser-machined out of two microfabricated, gold-sputtered alumina substrates.
A spacer separates the two trap layers.
The supply range of the dc segments of $\pm10$~V allow for axial trap frequencies of $2 \pi \times 0.2-4 $~MHz.
Typically, $160-600$ V$_{pp}$, at drive frequencies in the range $2\pi\times$~20-40~MHz, are applied to the RF electrodes, resulting in radial trap frequencies of $2\pi\times 2-4$~MHz. 
The fiber ends are shielded by the trap wafers from laser light, which enters through the trap slit perpendicular to the surface of Fig. \ref{fig:trap}.

\subsection{Ion shuttling and separation} 

A field programmable gate array based arbitrary wave form generator controls the voltages of the dc e\-lec\-tro\-des~\cite{Walther2012}.
It supplies output voltages in the $\pm$10~V range with a resolution of 0.3~mV and analog update rates up to $2.5$ MSamples/s, while having low noise ($\lesssim 10$~nV rms at trap frequencies).
Second-order \textPi-type low-pass filters for each segment suppress noise arising from voltage updates.

The segmented design allows for performing ion shuttling and ion separation.
We have demonstrated fast shuttling over $280$~\textmu m in $3.6$~\textmu s with an increase in motional quanta of only $0.10(1)$~\cite{Walther2012,Bowler2012}, and ion crystal separation operations, with an average increase of $\approx 4$ motional quanta per ion in $80$~\textmu s for a separation distance of 500~\textmu m~\cite{Kaufmann2014,Ruster2014}.


\subsection{Qubit preparation, manipulation and readout}
\label{sub:qubit}

\begin{figure}
		\includegraphics[width=.485\textwidth]{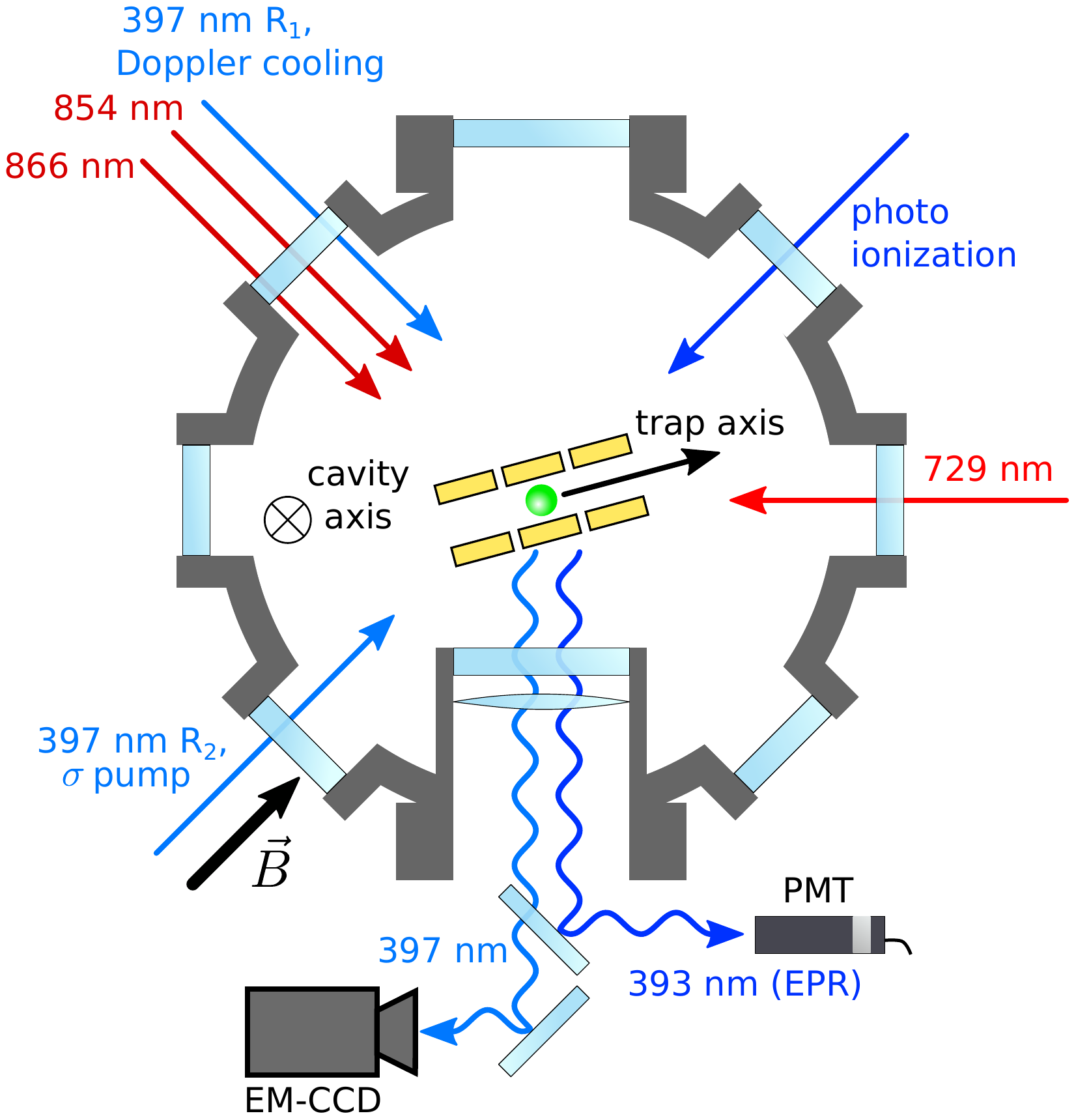}
	\caption{Beam geometry of the experimental setup, viewed from the top. 
	Arrows indicate directions for laser beam propagation and the quantization axis defined by the magnetic field $ \vec{B} $. 
	$ R_1 $ and $ R_2 $: Beams used for driving Raman transitions between $\ket{S_{1/2},\pm 1/2}$. 
	 Other lasers are described in the text.
	 A lens in the inverted viewport (bottom) collects scattered 397~nm light (and for the EPR-protocol, also 393~nm light) for detection. 
	 The cavity axis is perpendicular to plane of view.}
	\label{fig:chamber}       
\end{figure}

\begin{figure}
		\includegraphics[width=.48\textwidth]{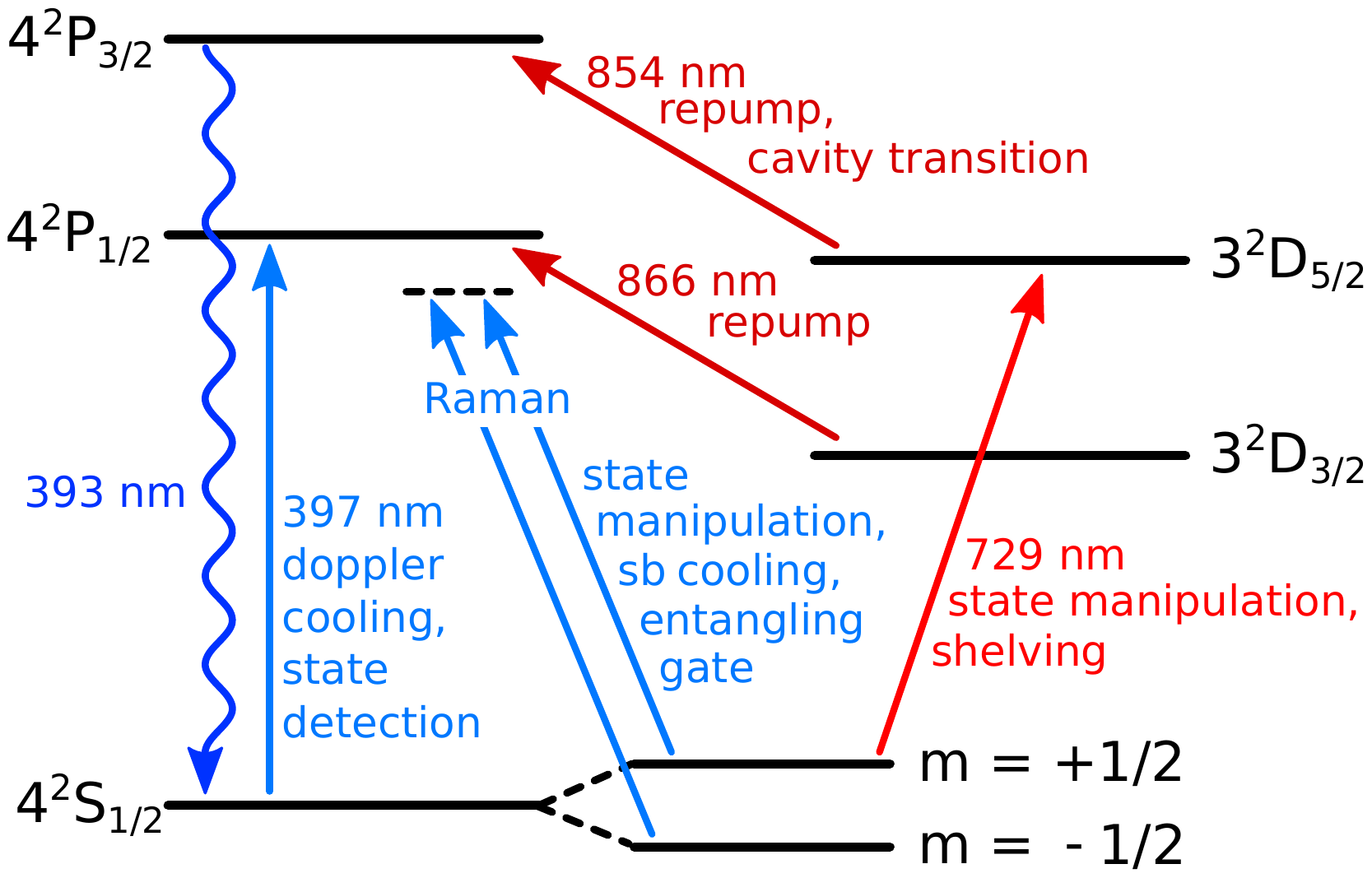}
	\caption{Level scheme of $^{40}$Ca$^+$ with all relevant transitions. For details see text.}
	\label{fig:lvl}       
\end{figure}

As the stationary qubit, we employ \Ca-ions, where all relevant electronic transitions can be driven with commercially available diode lasers. 
\Ca~allows for encoding a \textit{spin qubit}\cite{Poschinger2009}, where the ground state levels $\state{S}{1/2}{-1/2}$ and $\state{S}{1/2}{+1/2}$ represent the logical states.
Alternatively, we can utilize an \textit{optical qubit}\cite{1367-2630-15-12-123012}, where either one or two of the logical states is represented by one of the sublevels of the long-lived metastable D$_{5/2}$ state, see Fig. \ref{fig:lvl}. 
The coherence times for both qubits is in the range of  $10-100$~ms, limited by magnetic field fluctuation.
As both qubit types are employed for the quantum repeater schemes analyzed in the manuscript, we give a short explanation of how the  qubits are implemented.

\textbf{Ground state cooling:} 
For each experimental run, we start with Doppler cooling on the $S_{1/2} \leftrightarrow P_{1/2}$ cycling transition near 397~nm. 
We obtain a thermal state with a typical average phonon number of $\bar{n} \approx 20$ on the axial mode of vibration. 
We employ pulsed sideband cooling, by driving a stimulated Raman transition between the Zeeman ground states of the $S_{1/2}$ state, to cool close to the ground state of the axial mode. 
The repumping is accomplished by employing a circularly polarized laser field, driving the cycling transition. 
We typically attain average phonon numbers lower than 0.05 in the axial mode.

\textbf{Initialization:} 
The qubit can be initialized in the state $\ket{S_{1/2},-1/2}$ with high fidelity ($>0.99$) by repetitively transferring population from  $\ket{S_{1/2},1/2}$  to the $D_{5/2}$ state, and quenching the population back into the $S_{1/2}$ state by driving the $D_{5/2} \leftrightarrow P_{3/2}$ transition with a laser field near 854~nm. 
The population transfer in the first step is done by driving \textpi-pulses on a suitable subtransition $\ket{S_{1/2},1/2} \leftrightarrow \ket{D_{5/2},m_D}$, see Fig. \ref{fig:lvl}. 
The laser pulses are derived from a laser source near 729~nm, stabilized to a linewidth of below 1~kHz. 
The frequency is controlled using an acousto-optical modulator. 
The natural linewidth of the $\ket{S_{1/2}}\leftrightarrow\ket{D_{5/2}}$ quadrupole transition of $2 \pi \times 0.14$~Hz and the narrow laser linewidth allow for selectively driving transitions between different Zeeman sublevels.

\textbf{Coherent manipulation of single qubits:} 
Based on the preparation in  $\ket{S_{1/2},m_S}$, we can prepare arbitrary superposition states within $S_{1/2}$ and $D_{5/2}$ manifolds. 
Coherent rotations of the spin qubit are driven by stimulated Raman transitions.
At a Raman detuning of about $2\pi\times100$~GHz, we achieve \textpi-times of a few \textmu s.
Coherent rotations on the optical qubit are driven by a laser near 729~nm as explained for the initialization. 
Note that the coherent dynamics on the quadrupole transition depends on the motional state of the ions, such that we need to keep the ions in the Lamb-Dicke regime to achieve high-fidelity operations. 
By contrast, the rotations driven by radiofrequency or on the stimulated Raman transition are independent of the motional state.
An arbitrary spin qubit state can thus be mapped to the $D_{5/2}$ manifold by using a quadrupole-\textpi-pulse for each spin state. 
Note that coherent rotations between Zeeman sublevels of the same manifold can also be driven with radiofrequency pulses.

\textbf{State Readout:} 
Readout of the optical qubit is performed by fluorescence detection on the cycling transition with an EM-CCD camera or a photomulitplier tube (PMT).
A bright event corresponds to the $S_{1/2}$ state, and a dark event the $D_{5/2}$ state.
For the spin qubit, it is necessary to \emph{shelve} one of the qubit states by transferring population from it with a \textpi-pulse on the quadrupole transition.
The readout is then analogous to the case of the optical qubit.
In both cases, the readout fidelities of $\approx 0.995$ can be obtained.

\textbf{Entangling gate:} We entangle two stationary spin qubits by means of the \textit{geometric phase gate}~\cite{Leibfried2003a}, where spin-de\-pen\-dent dipole forces are employed to transiently excite motional modes depending on the spin configuration of an entire ion string. 
For spin configurations where a mode is displaced, a geometric phase is acquired. 
This conditional phase gives rise to entanglement. 
The spin-de\-pe\-ndent dipole force is created by employing the off-resonant laser beams which are also used for driving stimulated Raman transitions.  
We achieve Bell-state fidelities of up to 97\% (corrected for preparation and measurement errors) for gate durations of about 100~\textmu s.
The entangled spin qubits can be converted to Bell states of optical qubits by \textpi-pulses on the quadrupole transition, as explained for the coherent manipulation.

\textbf{Cavity-induced stimulated Raman transition:}
For the interaction between flying and stationary qubit, we can employ a cavity-induced stimulated Raman transition between the $S_{1/2}$ and $D_{5/2}$ states, driven by a laser off-resonant to the cycling transition and the cavity field.
In this case, the coupling strength is given by
\begin{equation}
	\Omega^\text{eff}=\frac{G\, g\,\Omega_L}{\Delta},
\end{equation}
where $\Omega_L$ is the on-resonance Rabi frequency of the laser, $g$ the cavity vacuum coupling rate and $\Delta$ the laser detuning.
$G = cg \cdot \mathcal{P_\mathbf{d}}(\mathbf{\boldsymbol\epsilon})$ combines the Clebsch-Gordan coefficient $cg$ for both transitions with the projections $\mathcal{P_\mathbf{d}}(\mathbf{\boldsymbol\epsilon})$ of the polarization $\mathbf{\boldsymbol\epsilon}$ of the laser and the cavity field onto the ionic dipole moment $\mathbf{d}$, see Fig. \ref{fig:chamber}.
We estimate, based the cavity properties (Sec. \ref{sub:cavity}) and the available laser power, that effective transition frequencies of $\Omega^\text{eff} \approx 1$~MHz are within reach.

\subsection{Fiber based cavity}
\label{sub:cavity}

In order to achieve a large coupling of the electronic state of an ion with the cavity mode, the mode volume is kept as small as possible.
Due to its small size, a fiber based Fabry-P\'erot-cavity\cite{Colombe2007,Steiner2014,Brandstaetter2013}, where highly reflective dielectric mirrors are sputtered on end facets of optical fibers, can fulfill this requirement.
It is suited to be accommodated between the two electrode chips, providing direct coupling into the cavity via one of the fibers.

The cavity drives the $D_{5/2}\leftrightarrow P_{3/2}$ transition near $\lambda=$854~nm. 
The respective field coupling parameter, i.e., the vacuum Rabi frequency $g_0$, is given by:

\begin{equation}\label{eq_g0}
g_0= \sqrt{\frac{3 \, c \lambda^2 \gamma_\text{\tiny{PD}}}{\pi^2w_0^2L}}
\end{equation}
where $ \gamma_\text{\tiny{PD}} = 2 \pi \times 0.67 $~MHz is the radiative field decay rate of the $D_{5/2} \leftrightarrow P_{3/2}$ transition. 
The cavity mode waist $w_0 $ is set by choosing the radius of curvature (ROC) and the length $ L $ of the cavity.
The range of suitable values for $ L $ is predetermined by the trap dimensions:
Both the fibers and the mirror surfaces are comprised of insulating materials, which are prone to uncontrolled charging when exposed to UV laser light, leading to uncontrolled electric stray fields in the trap\cite{Harlander2010,Herskind2011}.
However, the trap dimensions cannot be arbitrarily small, since short distances of the electrode surfaces to the ion increase anomalous heating rates\cite{Brownnutt2014} and the optical access needs to be ensured.
The cavity length $ L $ is thus chosen sufficiently large, such that the fibers are retracted behind the trap electrodes (see Fig. \ref{fig:trap}), reducing the electrical feedthrough of the charged insulating surfaces to the trap volume.
We utilize a plano-concave cavity setup to reach a high mode matching $ \varepsilon $ between the mode that emanates from the fiber and the cavity mode.
The ROC of the concave mirror can be chosen such that the waist is small while cavity stability and high mode matching are ensured.

In our case, the electrodes are separated by $ 250$~\textmu m (see Fig. \ref{fig:trap}), which demands a large diameter concave mirror structure on the fiber facet to avoid finesse limitations by clipping losses.
We developed a novel technique for shaping these facets using a commercial focused ion beam (FIB) device\footnote{\textit{FEI Helios NanoLab}}, which allows us to create spherical structures with a large range of possible ROCs\footnote{M. Salz, \textit{to be published}}.
Our cavity setup has a length of $L= 250~$\textmu m and consists of a singlemode fiber with a plane surface and a multimode fiber with $ 350$~\textmu m ROC concave facet.
The facets are coated with dielectric mirror layers with a target transmission of 50(15)~ppm at a wavelength of 854~nm\footnote{\textit{Laser Optik Garbsen}}.
The linewidth was determined to $2\,\kappa= 2 \pi \times 36.6(5)~$MHz using frequency modulation as a frequency marker.
The field decay rate of the cavity follows as $ \kappa = 2 \pi \times 18.3 (3) $~MHz\,. 
A finesse of $\mathcal{F}=1.65(2)\times 10^4$ is deduced.
The cavity has a mode waist of $w_0=6.6$~\textmu m, i.e., the maximum cavity-ion coupling parameter at the plane mirror is $ g_0= 2\pi\times25.7~\text{MHz} $.
In the cavity center, the coupling is reduced to 
\begin{equation}\label{eq_gc}
g_c=\frac{w_0}{w(L/2)}g_0 
\end{equation}

The set of cavity parameters thus reads $ (g_c,\kappa,\gamma_\text{\tiny{PD}},\gamma_\text{\tiny{PS}})$ $=2\pi\times(20.1,18.3,0.67,10.7)~$MHz, which means that the cavity operates in the intermediate coupling regime.
As the decay rate $P_{3/2}\rightarrow S_{1/2}$ is as strong as the coupling parameter, this system will not display resonant coherent dynamics on the $D_{5/2} \leftrightarrow P_{3/2}$ transition.
However, excitation and off-resonant dynamics supported by the cavity can be utilized.
Further effective reduction of the cavity coupling $ g_c $ due to geometrical considerations or Clebsch-Gordan coefficients of a particular atomic transition will be taken into account in the discussion of the protocol efficiencies in Sec.~\ref{sec:assess}.

It has been recently shown \cite{Gallego2015} Mode matching for a fiber based cavity differs from the usual approach for Fabry-P\'erot cavities as found, e.g. in \cite{meschede_book2005}.
Most importantly, the minimum of the reflection signal no longer corresponds to the optimal incoupling.
However, in our case we find minimal corrections, and will use the latter approach for brevity here.
When coupling light into the cavity, two effects reduce the contrast of the reflection dip on resonance $\eta_\text{dip}= \varepsilon\cdot\eta_{\text{imp}}$\cite{Hood2001,Gallego2015}: The mode matching $ \varepsilon $ and the impedance matching, described by the coefficient $ \eta_{\text{imp}} $, which depends on the transmission $ T $ of the cavity mirrors and the total losses $ \mathcal{L} $ per round trip.
The losses are determined to be $ \mathcal{L}=280(30)~$ppm. For symmetric coating $ T_1=T_2\equiv T $:

\begin{equation}\label{eq_eta-impedance}
\eta_{\text{imp}}=1-\left(\frac{ \mathcal{L}}{2\,T+ \mathcal{L}}\right)^2= (45.6\pm9.1)\%
\end{equation}

Comparing this to the experimentally observed contrast of $ \eta_\text{dip}=20.3(1)\% $, one can find

\begin{equation}\label{eq_epsilon}
\varepsilon=\frac{\eta_\text{dip}}{\eta_{\text{imp}}}=(44.5\pm 8.9)\%
\end{equation}

The probability that a resonant photon emitted from the ion enters the cavity mode is enhanced by the Purcell effect and given by
\begin{equation}
\label{eq:eta_P}
	\eta_P = \frac{2 C_c}{2 C_c + 1}=0.97
\end{equation}
where $C_c$ is the cooperativity in the cavity center,
\begin{equation}
\label{eq:C_c}
	C_c = \frac{g_c^2}{2 \, \gamma_\text{\tiny{PD}} \, \kappa}=16.5~.
\end{equation}
Neglecting mode matching, such a photon then has the probability
\begin{equation}
\label{eq:eta_CT}
	\eta_{out} = \frac{T}{2T + \mathcal{L}} = 0.13
\end{equation}
of leaving the cavity through the single mode mirror, which is the ratio of this mirror's transmission loss to the sum of all loss channels of the cavity.

To ensure the frequency stability of the fiber cavity, the fiber cavity can be actively stabilized to a laser before each experimental shot.

The setup so far features only one input/output (I/O) port, in the form of the fiber-based cavity.
In order to extend the setup to actual QR chains, the singlemode I/O fiber can also be equipped with a fiber switching device (i.e., before the left fiber of Fig. \ref{fig:sequence}).

\section{Assessment of the possible protocol implementations}
\label{sec:assess}

\begin{figure*}
	\centering
		\includegraphics[width=.95\textwidth]{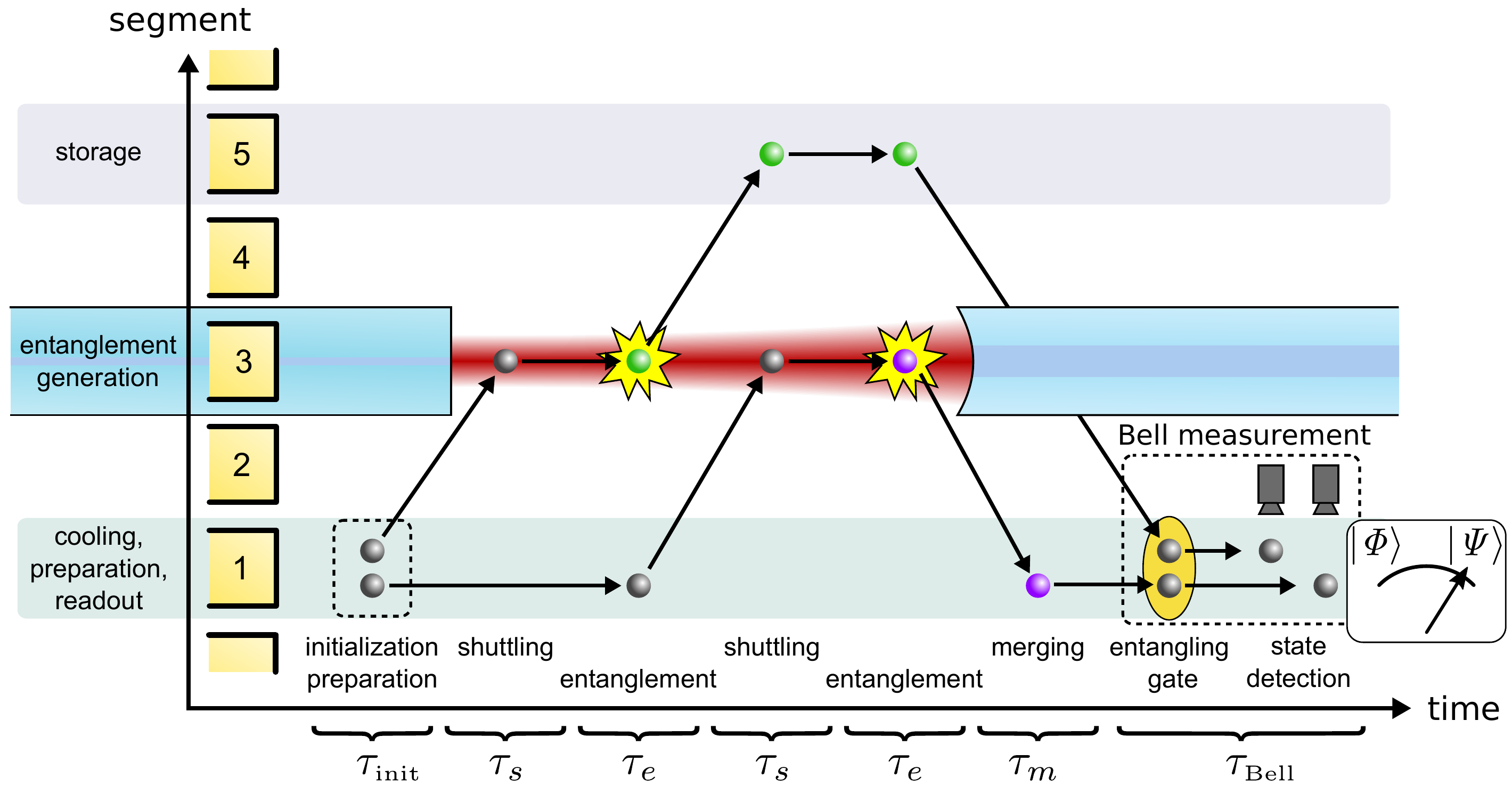}
	\caption{Experimental sequence for a basic quantum repeating operation with the presented setup. 
	The trap axis is divided into three regions with different purposes: 
	ions are shuttled into the region around segment 1 for cooling, state preparation, manipulation and readout. 
	The ions are shuttled into the region around electrode 3, into the fiber cavity, for entanglement generation. 
	The region around electrode 5 serves as a short-term storage.
	After cooling and initializing both (unentangled, grey) ions, one is shuttled into the cavity where entanglement with a qubit from QR (N$-$1) is generated by either of the three protocols (entanglement indicated by color), see sections \ref{sub:exEPR}- \ref{sub:expHyb}.
	 After shuttling the first ion to the short-time storage, the second ion is shuttled into the cavity and the same operation is performed with a qubit from QR (N+1). 
	 This requires a fiber switch in front of the single mode fiber (left), see Sec. \ref{sub:cavity}.
	 The ions are then shuttled back to segment 1, a two-ion crystal is formed and a Bell state measurement is performed. 
}
	\label{fig:sequence}       
\end{figure*}

In this section we investigate how the different repeater protocols can be implemented on our hardware platform. 
We specify the experimenal requirements and derive possible experimental sequences. 
For each protocol, we quantitatively estimate the attainable fidelities and entanglement generation rates for spatially separated Bell states at QR (N) and QR (N+1).
Fig. \ref{fig:sequence} depicts the experimental sequence at QR (N) in the chain of repeater nodes (see Fig. \ref{fig:QrepPrinciple}). The neighboring QRs have an analogous sequence.

All protocols have similar initialization sequences in the beginning and entanglement swapping sequences in the end, see Sec. \ref{sub:qubit}.
The duration of the entire protocol run in combination with the success probability determines the entanglement generation rate of the protocol.
The time necessary to run one entangling sequence, from initialization to entanglement of a local ion with a distant one, is called $\tau_\text{run}$, and is split into a preparation time $\tau_\text{prep}$, and $\tau_e$.
The latter is the time each protocol requires to entangle the two distant ions, once they are initialized and in the cavity.
The time $\tau_\text{prep}\approx 210$~\textmu s includes the initialization of the ion state $\tau_\text{init} \approx 10$~\textmu s, the shuttling of the ion into the cavity $\tau_S \approx 100$~\textmu s, as well as the shuttling back to the processor region to be re-initialized if the protocol does not succeed, which again takes the time $\tau_S$. 
It does not include the post-processing of the ions necessary after the successful entanglement has been heralded.
Errors \textit{before} the herald detection decrease the success probability, whereas errors obtained \textit{after} the herald detection reduce the fidelity $F$.

\subsection{Distributed EPR protocol}
\label{sub:exEPR}

\begin{figure}
	\centering
		\includegraphics[width=.45\textwidth]{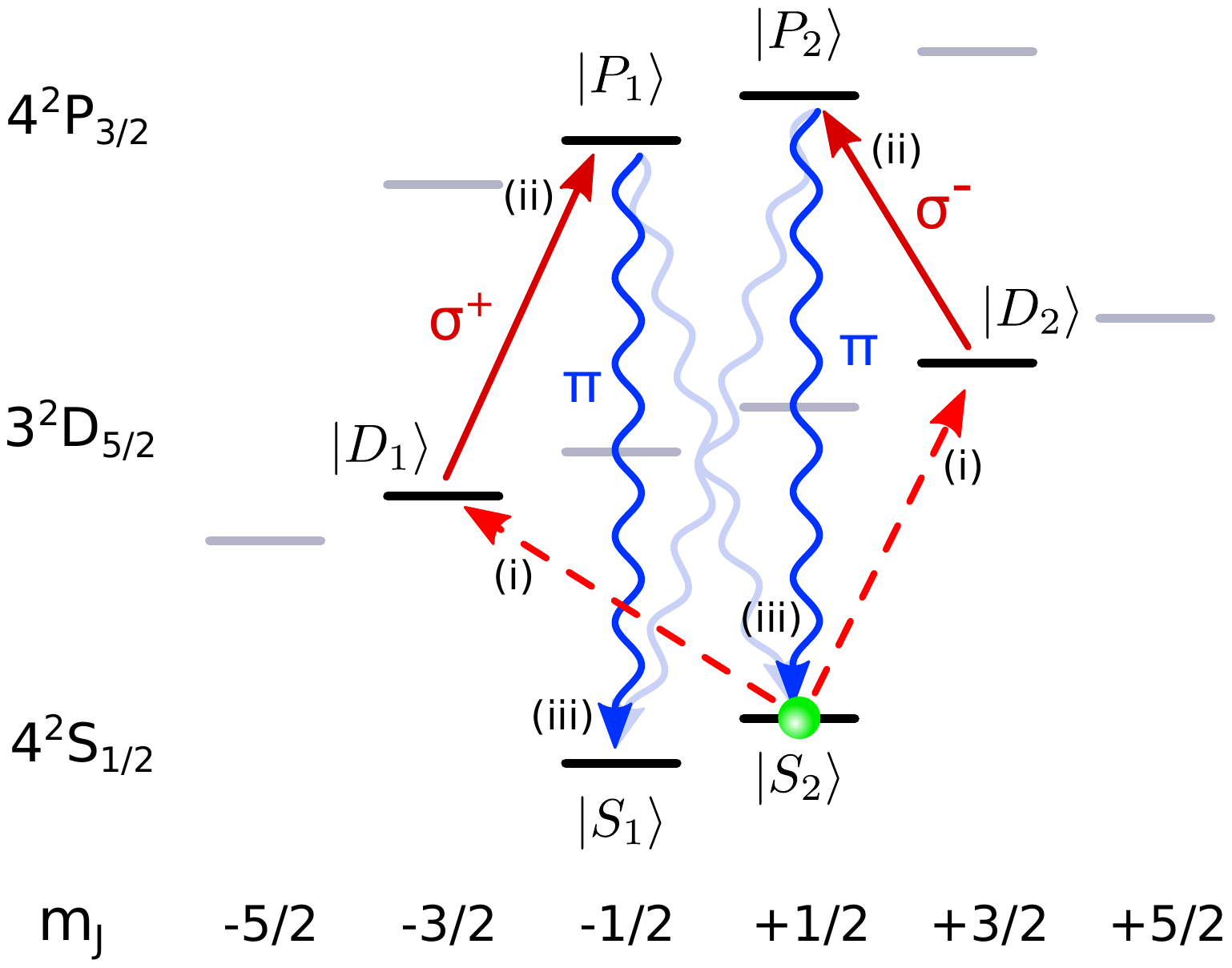}
	\caption{Level scheme of a $^{40}$Ca$^+$-ion with relevant transitions for the distributed EPR protocol. 
Encircled numbers indicate protocol steps. 
(i) After cooling and pumping to the $ \ket{S_2} $ state, the ion is initialized in a superposition of $ \ket{D_1} $ and $ \ket{D_2} $ (dashed red arrows).
(ii) In the cavity, the entangled photon induces a transition to one of the $ \ket{P} $-states, depending on its polarization, $ \sigma^+ $ or $ \sigma^- $ (dark red arrows).
(iii) The ion quickly decays into one of the $ \ket{S} $-states under emission of a photon at 393~nm as herald. 
By filtering out circular polarizations (light blue waves), only the events which preserve the information (dark blue wave) are detected.}
	\label{fig:eprlvl}       
\end{figure}

To implement the EPR protocol in our apparatus, we finish the initialization process from Sec. \ref{sub:qubit} by creating a superposition between the states $\state{D}{5/2}{-3/2}$ and $\state{D}{5/2}{+3/2}$, in the following denoted as $\ket{D_1}$ and $\ket{D_2}$, using suitable pulses on the $S \leftrightarrow D$ quadrupole transition (see Fig. \ref{fig:eprlvl}). 



We thus choose the starting state of each stationary qubit as (Fig. \ref{fig:eprlvl}(i))
\begin{equation}
	\ket{\Psi_D(t)}=\ket{D_1}+e^{i\phi_D(t)}\ket{D_2},
\end{equation}
with the phase
\begin{equation}
\label{eq:epr:phase}
\phi_D(t)=(\Delta m_D g_D - \Delta m_S g_S)\frac{\mu_B}{\hbar}B t + \phi_0.
\end{equation}
The $\Delta m_i$ are the differences of the magnetic quantum numbers of the respective manifold, the $g_i$ are their  Land\'e factors, $\mu_B$ is the Bohr magneton, and $\phi_0$ the phase offset at initialization.

After transporting the ion into the cavity, an EPR source between QR (N) and QR (N+1) provides the polarization entangled photon pairs as flying qubits for this repeater protocol, producing the state
\begin{align}
	\ket{\Psi_\text{EPR}} = \ket{L}\ket{R}+\ket{R} \ket{L}
\end{align}
with left-handed (L) and right-handed (R) circularly polarized light.
To ensure that the frequency and linewidth of the photon match the $D_{5/2} \leftrightarrow P_{3/2}$ transition close to $854$~nm, one use a Single Parametric Downconversion (SPDC) source with a filter cavity~\cite{Schuck2010}.
The production rate of entangled photon pairs after cavity filtering is denoted $r_\text{EPR}$.

The photons are coupled into the fibers whose ends constitute the cavities of QR (N) and QR (N+1) with an efficiency $\eta_{FC}$, and from there into the cavities with an incoupling efficiency between fiber and cavity  $\eta_\text{dip}$.
A photon successfully injected into the cavity of the node interacts with the ion (Fig. \ref{fig:eprlvl}(ii)) with a probability $P_{\text{int}}$, giving rise to the following state evolution:

\begin{align}
\label{eq:eprtrans}
&\ket{\Psi_D} \times \ket{\Psi_\text{EPR}} & \\
	\xrightarrow{abs.}& \ket{P_1}+ e^{i \phi_D}  \ket{P_2} & \nonumber \\ 
	\xrightarrow{em.}&  \big(\ket{S_1} + e^{i \phi_D}  \ket{S_2}\big)\ket{\pi_{393}} \nonumber
\end{align}
$\phi_D$ is taken from Eq. \ref{eq:epr:phase}, at the time $t$ when the ion absorbs the photon and decays into $S$.

In the case of success, the interaction maps the state of one photon onto the two Zeeman sublevels of the ion's $S_{1/2}$ state, under creation of a \textpi-polarized herald photon at a wavelength near $393$~nm (Fig. \ref{fig:eprlvl}(iii)), detected with probability $P_\text{det}$.
A detection event in each node's herald detector, within a detection window $\tau_c$ of each other, denotes a successful entanglement of ions in QR (N) and QR (N+1), after which the ion is moved out of the fiber cavity, and the second ion is entangled with an ion at QR (N$-$1).

We assume a rate of EPR photon pairs resonant with the $D_{5/2}\leftrightarrow P_{3/2}$ transition is $r_\text{EPR}=7800$~s$^{-1}$~\cite{Huwer2013}, after the filter cavity.
Throughout this section, it is assumed that only one photon ever populates the cavity mode at the same mode.
This approximation breaks down as soon as the cavity population decay rate $2 \, \kappa$ is no longer considerably greater than the EPR arrival rate at a single cavity, $\kappa\not\gg r_{\text{EPR}}\,\eta_{FC}\,\varepsilon$.
With $\kappa = 2 \pi \times 18.3$~MHz, such rates of EPR pair production are beyond current technological proficiency, and our approximation holds.

Solving the Liouville master equation for our setup, including all sublevels of the $S$, $P$ and $D$ states, with cavity parameters as given in Sec. \ref{sub:cavity}, returns the interaction probability of $P_\text{int} = 0.047$\,.
Required is a Zeeman splitting less than the cavity bandwidth $2 \, \kappa$, thus limiting the magnetic field to $B < 11$~G in order to ensure that the cavity field drives the transitions $D_1\rightarrow P_1$ and $D_2 \rightarrow P_2$ at similar rates.

The $\eta$ in Eqs. \ref{eq:eprr1} and \ref{eq:eprr2}, the probability of one of the photons after the filter cavity to interact with one ion, is given by
\begin{gather}
	\eta = \eta_{FC}\, \eta_\text{in} \, P_\text{int}\approx 0.005 \\ 
	\eta^2 \approx 3 \times 10^{-5} \,.\nonumber
\end{gather}
Here, $\eta_{FC} \approx 0.9$ is the efficiency of coupling the photons into the fiber after the filter cavity, and \mbox{$\eta_\text{in} = \eta_\text{out} = 0.13$} is the probability of coupling a single photon into the cavity, by time-reversal symmetry to the outcoupling process.

The detection probability 
\begin{equation}
	P_\text{det} = \frac{d\Omega}{4\pi}\, \eta_{QE}^{397} = 0.007
\end{equation}
depends of the solid angle of our detection lens $d\Omega/4\pi$, and the quantum efficiency $\eta_{QE}^{397}$ of the PMT for UV light (Sec. \ref{sec:setup}).


\textbf{Entanglement generation Rate:}
With these efficiencies, the entanglement generation rate, Eq. \ref{eq:eprRate}, takes on the shape
\begin{align}
\label{eq:eprre}
	r_e(\tau_W) =& \frac{P_e(\tau_W)}{\tau_\text{prep} + \tau_W}
\end{align}
for a given time to wait on an entanglement event $\tau_W$.

We have yet to account for dark counts on the detectors, which both increase the apparent rate of successful events and decrease the fidelity of our final state.
A typical dark count rate for UV detectors can be assumed to be $r_\text{dc}=60$~s$^{-1}$~\cite{Kurz2015}.
For a detection window $\tau_c$, during which detector events are counted as concurrent, the probability of registering two simultaneous dark counts is given by
\begin{equation}
\label{eq:2dc}
	P_{2dc}(t,\tau_c) = (1-e^{-r_\text{dc} t})\cdot(1-e^{-r_\text{dc} \tau_c})\,.
\end{equation}
Similarly, a photon-qubit interaction in one of the nodes can be concurrent with a dark count in the other, with a probability of
\begin{equation}
\label{eq:hdc}
	P_{hdc}(t,\tau_c)= P_\text{det} P_1(t)\cdot(1-e^{-r_\text{dc} \tau_c}) \,,
\end{equation}
with $P_1(t)$ from Eq. \ref{eq:eprP1} in Sec. \ref{EPR}.

The probability $P'_e(t,\tau_c)$ of any two-detector event, true or false positive, happening during time $t$, where the window for coincidenct detection is set to $\tau_c$, is given by
\begin{gather}
\label{eq:EPRRate}
	P'_e(t,\tau_c) = P_e(t) + P_{2dc}(t,\tau_c) + P_{hdc}(t,\tau_c) \\
	\Rightarrow r'_e=\frac{P'_e(t,\tau_c)}{\tau_\text{prep} + \tau_W +\tau_c}\,.
\end{gather}

We set $\tau_W=r_1^{-1} \approx 15$~ms (see  Eq. \ref{eq:eprr1}) to the time by which we can expect one of the photons to have interacted with an ion, making a new initialization necessary.
Using $\tau_\text{prep}\approx 210$~\textmu s (Sec. \ref{sec:assess}.0), and $\tau_c =\frac{1}{20 \kappa}$, the repetition rate becomes $r_e(\tau_W) = 6.4 \times 10^{-6}$.
Inserting the values of the setup, this culminates in a rate $r'_e \approx 7.4 \times 10^{-6} $~s$^{-1}$, i.e. about one event every 35 hours.
We see that a highly efficient detection of herald photons at 397~nm together with an effective in-coupling of EPR photons into the cavity is required to improve this rate. 

\textbf{Fidelity:}
The fidelity of the final state is limited by the ratio of real entanglement events to total heralded events, $F = r_e/r'_e \approx 0.86$. 
Another fidelity error is the imperfect filtering of \textsigma-light for detection optics that is not pointlike in extent.
For a solid angle of $d\Omega= 4\pi \cdot 0.035$ of the collecting lens, the opening angle is $\theta=21^\circ$.
However, integrating the arriving herald wave function over this area shows the amount of \textsigma-light in the wrong mode to be negligible.

Finally, imperfect initialization and readout (Sec. \ref{sub:qubit}) reduce the fidelity to $F \approx 0.86 \cdot 0.995$ for the EPR protocol.


\subsection{DLCZ protocol}
\label{sub:exDLCZ}

\begin{figure}
	\centering
		\includegraphics[width=.4\textwidth]{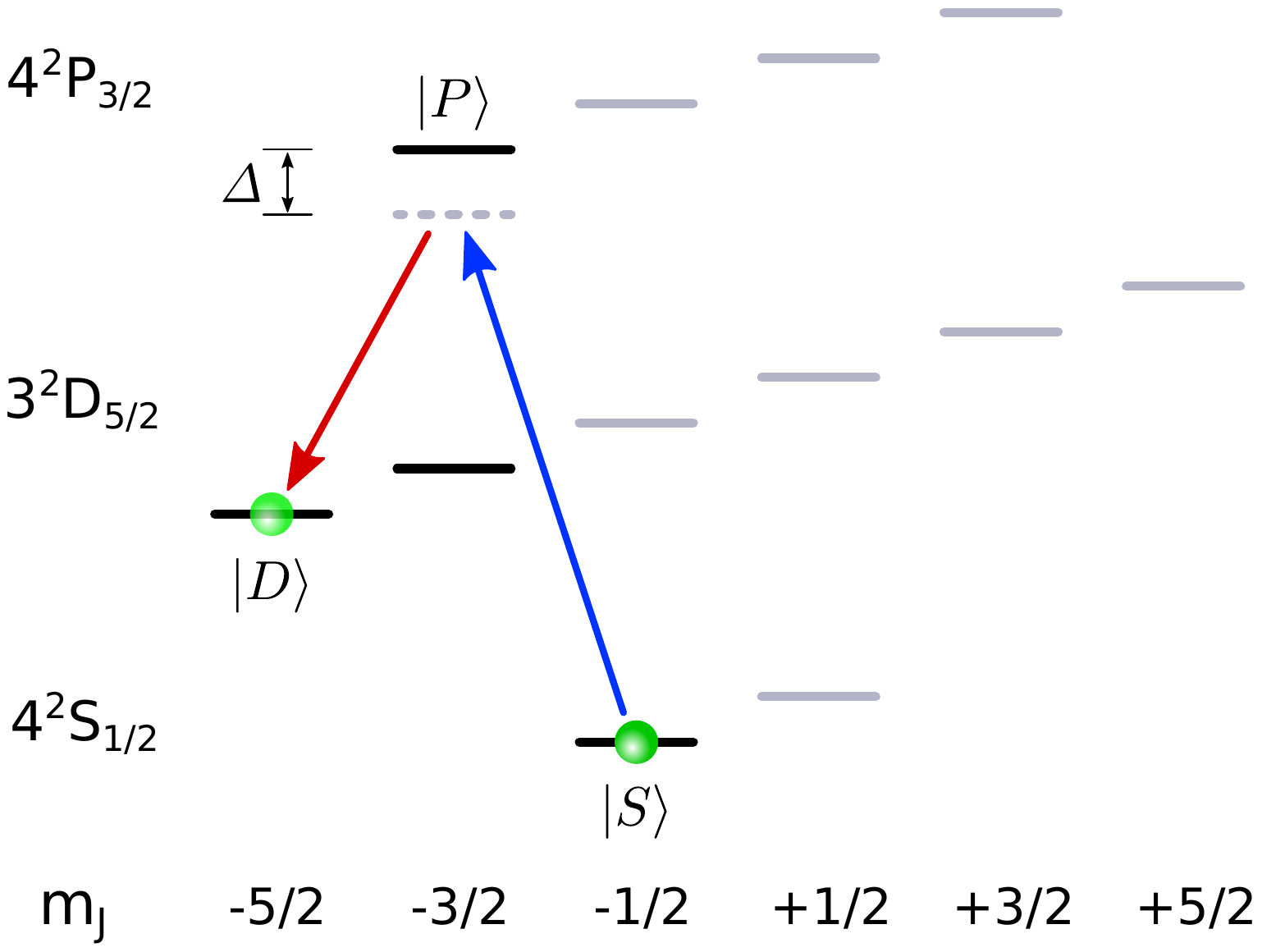}
	\caption{Level scheme with relevant transitions for the single-photon DLCZ protocol. 
After initialization in the $ \ket{S} $-state, a cavity-induced stimulated Raman transition is driven to $ \ket{D} $ via the $ \ket{P} $ state, consisting of a Raman pulse (blue arrow) with detuning $ \Delta $, and the cavity fulfilling the Raman resonance condition with the beam. 
A single cavity photon is generated for both nodes, and the detection setup eliminates which-way information (see Fig. \ref{fig:theo} b) to entangle the nodes in $\ket{S,D}+\ket{D,S}$.}
	\label{fig:dlczlvl1}       
\end{figure}

The implementation of the DLCZ protocol in our apparatus, depicted in Fig. \ref{fig:dlczlvl1}, starts with initializing the ion in the state

\begin{equation}
	\ket{S} = \state{S}{1/2}{-1/2}.
\end{equation}

After transport of the ion into the cavity, a photon is created by a cavity-induced stimulated Raman transition (Sec. \ref{sub:qubit}) between the $S_{1/2}$ and the $D_{5/2}$ manifolds, with $\ket{0} = \ket{D}$ and $\ket{1} = \ket{S}$ chosen as shown in Fig. \ref{fig:dlczlvl1}.

The photon is emitted into the cavity with a probability $p_1$, which is controlled by the duration of the Raman drive.
It is detected with a probability $P_\text{det}$ once emitted.
This probability is given by
\begin{equation}
\label{eq:dlczdet}
	P_\text{det} = \eta_{P} \cdot \eta_\text{out} \cdot \varepsilon \cdot \eta_{QE}^{854}\,,
\end{equation} 
where $\eta_P$ is the probability that the photon is emitted into the cavity mode through Purcell enhancement, $\eta_\text{out}$ is the the cavity outcoupling coefficient,  $\varepsilon$ the mode matching efficiency between cavity and fiber and $\eta_{QE}^{854} = 0.5$ the quantum efficiency the photon detectors near $ 854$~nm.
All of these parameters are defined and given in Sec. \ref{sec:setup}, and lead to a detection efficiency of about $P_\text{det}=0.03$.

For a threshold fidelity of $F_\text{thr} = 0.99$, and thus a single photon emission probability $p_1 = 0.01$, these parameters result in a success probability for one experimental run of $P_1 \approx 6 \times 10^{-4}$ according to Eq. \ref{eq:dlcz1suc}.

\textbf{Entanglement generation rate:}
Each single experimental run can be done in time $\tau_\text{run}=\tau_\text{prep}+\tau_{R}+\tau_{\kappa}$, which is the sum of the preparation time $\tau_\text{prep}$, the Raman pulse time $\tau_R$, and the cavity decay time $\tau_\kappa$ (negligible in our case).
For a threshold fidelity of $F_\text{thr} = 0.99$, the single photon emission probability must fulfill $p_1 \le 0.01$ according to Eq. \ref{eq:dlcz:1suc}.
A typical duration of the Raman drive pulse is $\tau_R = 2$~\textmu s.
The preparation time $\tau_\text{prep}\approx 210$~\textmu s is substantially longer, leading to $\tau_\text{run} \approx 212$~\textmu s.
The entanglement generation rate for these parameters results in $r_e \approx 2.8$~s$^{-1}$. 

\textbf{Fidelity:}
The fidelity loss due to events where two photons are generated from driving the stimulated Raman transition is preset by choice of $F_\text{thr}$ and corresponding tuning of the drive pulse area. Additional fidelity loss is due to dark count events. 
However, choosing a detection window of $\tau_\text{det}=2\tau_\kappa$ after the Raman pulse in order to capture the majority of real events, with typical dark count rates of $20$~Hz for IR detectors, erroneous events are 5 orders of magnitude less frequent than entangling events.

Imperfect initialization and readout (Sec. \ref{sub:qubit}) also reduce the fidelity by a factor of $F_\text{init} \approx 0.995$, similar to the EPR protocol, so the final fidelity for $F_\text{thr}=0.99$ is $F_1\approx 0.99 \cdot 0.995$ .

\textbf{Two-photon-detection DLCZ:}
We also investigate an alternative version of the DLCZ protocol based on deterministic photon emission and coincident two-photon detection \cite{Simon2003,Barros2009}.

\begin{figure}
	\centering
		\includegraphics[width=.4\textwidth]{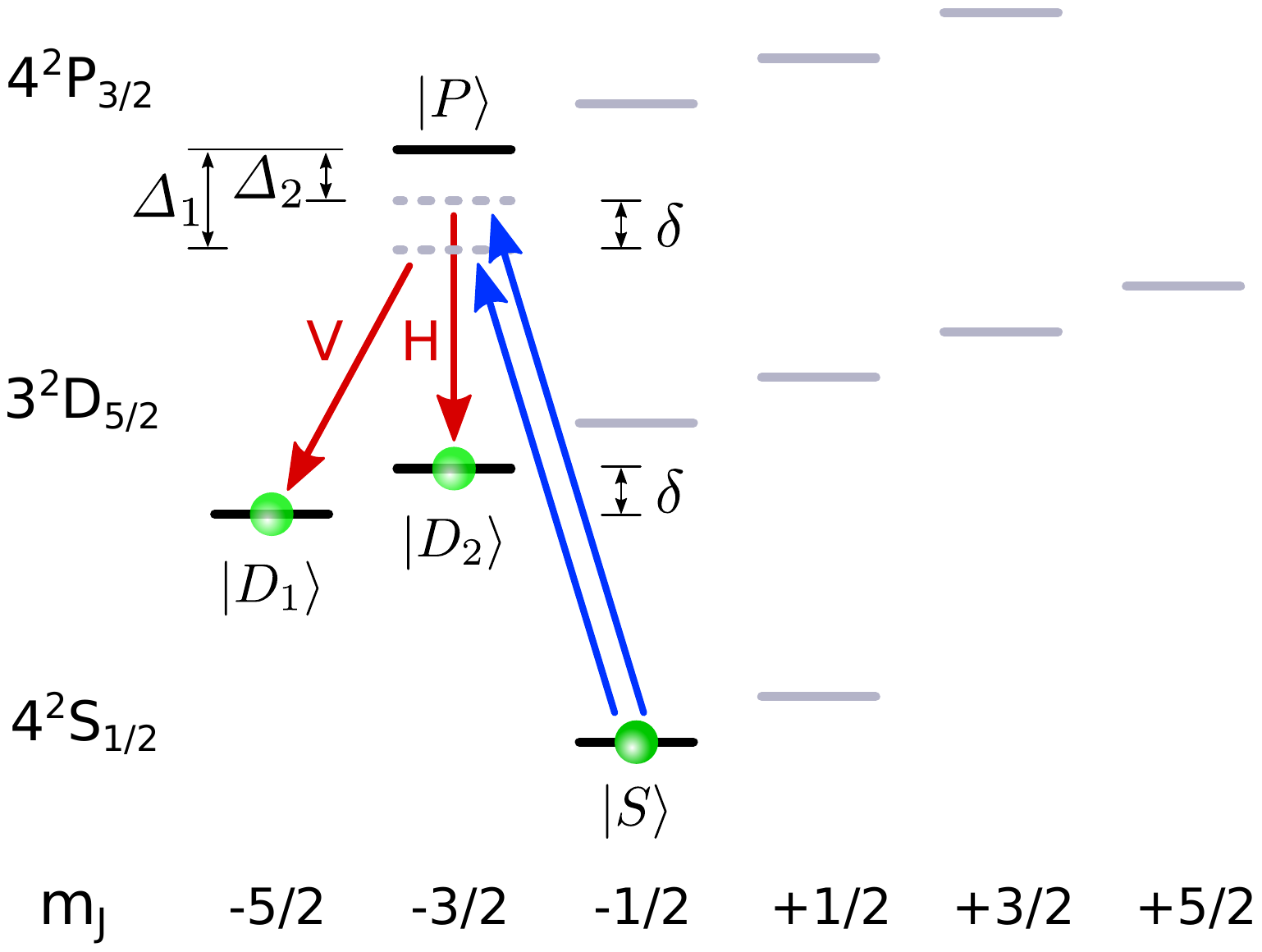}
	\caption{Level scheme with relevant transitions for the two-photon DLCZ protocol following~\cite{Stute2012}. 
After initialization in the $ \ket{S} $-state, two Raman transitions are driven simultaneously to $ \ket{D_1} $ and $ \ket{D_2} $ via the $ \ket{P} $ state, consisting of a bichromatic Raman pulse (blue arrows) with detunings $ \Delta_1 $, $ \Delta_2 $ and two modes of the cavity ($ V $ and $ H $ with respect to the cavity axis, red arrows). 
A cavity photon is generated, whose polarization $ V $ or $ H $ is entangled to the electronic state of the ion, $ \ket{D_1} $ or $ \ket{D_2} $, respectively.}
	\label{fig:dlczlvl2}       
\end{figure}

\begin{figure}
	\centering
		\includegraphics[width=.4\textwidth]{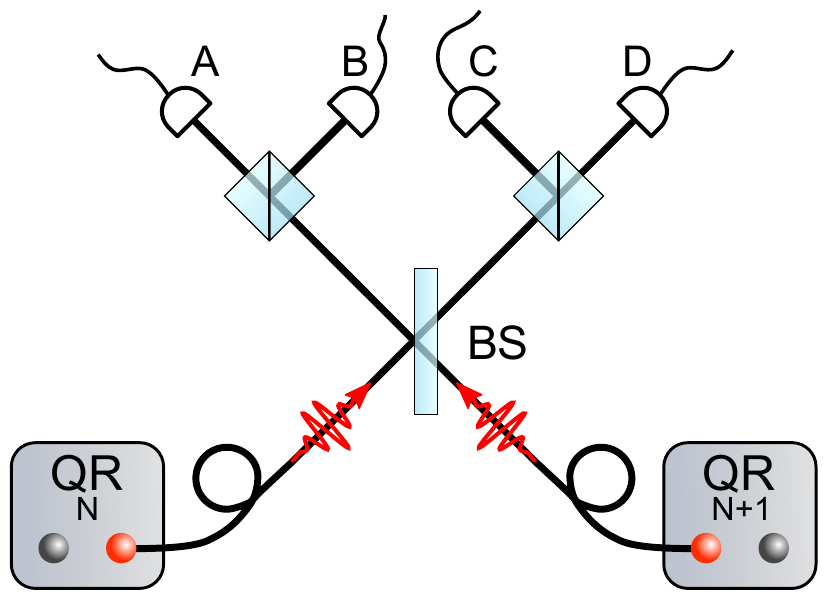}
	\caption{Alternative to the DLCZ method introduced in Sec. \ref{DLCZ}.
	Both nodes emit a photon induced by a Raman transition, which is entangled with the ion according to Eq. \ref{eq:dlczRaman2}. The photons are brought to interference within a detection setup as pictured. Coincident clicks in detectors \{AB, CD, AC, BD\} project the two ions onto a Bell state.
	}
	\label{fig:dlczDet2}       
\end{figure}

The single Raman beam of Sec. \ref{DLCZ} is substituted by a bichromatic Raman beam, so that two cavity-induced stimulated Raman transitions are driven, see Fig. \ref{fig:dlczlvl2}:
\begin{equation}
\label{eq:dlczRaman2}
	\ket{S}\ket{0}_p\xrightarrow{\text{Raman}}\ket{D_1} \ket{V} + \ket{D_2} \ket{H}\,.
\end{equation}
Note that $\ket{D_1}$ and $\ket{D_2}$ differ from the levels defined in Sec. \ref{sub:exEPR}.
The photons emitted on the \textsigma- and \textpi-transitions  are mapped to the $\ket{V}$ and the $\ket{H}$ mode of the cavity by setting the quantization axis of the ion to right angles with the cavity axis.

The stationary qubits are entangled by two-photon detection.
As~\cite{Zippilli2008} elaborates, this requires a 50/50 beamsplitter, with a polarizing beamsplitter (PBS) and two detectors at each output port, as depicted in Fig. \ref{fig:dlczDet2}.
Coincident clicks in detectors \{AB, CD, AC, BD\} project the two-qubit state onto the Bell state \{$\phi^+$, $\phi^-$, $\psi^+$, $\psi^-$\}.
As these possibilities represent half of the two-photon detection events possible, this protocol cannot exceed a success probability of $P_2 = \frac{1}{2}$.
Including the probability of detection $P_\text{det}$ for each photon, and the probability $p_1$ of successfully inducing the Raman transition from Eq. \ref{eq:dlczRaman2}, which can be set as close to unity as possible, the success probability of this protocol is
\begin{equation}
\label{eq:dlcz2suc}
	P_2 = \frac{1}{2} p_1^2 P_\text{det}^2 \approx 4 \times 10^{-4}\,.
\end{equation}

The two-photon detection eliminates the fidelity's dependence on photon loss, and is limited by the initialization and readout losses ($F=0.995$), and the width of the cavity compared to the Zeeman splitting of the used transitions.
Assuming a cavity centered on the transitions shown in Fig. \ref{fig:dlczlvl2}, a magnetic field of $10$~G, and taking into account the Clebsch-Gordan-Coefficients for the transitions, the parasitic transition $\ket{S}\rightarrow \ket{D_{5/2},-1/2}$ has an excitation probability of $\approx0.75$~\%.
These detrimental effects reduce the fidelity to $F = 0.992$\,.
Due to the coincident detection scheme, typical dark count rates for modern detectors lead to negligible errors.

The single experiment run time of this protocol is slightly lower than in the probabilistic DLCZ case, $\tau_\text{run} \approx 240$~\textmu s, as the Raman pulse duration is longer, $\tau_R \approx 30$~\textmu s, in order to achieve a \textpi-pulse.
The entanglement generation rate follows as $r_e \approx 1.6$~s$^{-1}$.

This is slightly higher than the respective rate for the single-photon-detection DLCZ protocol at $F=0.99~.$
Additionally, the two-photon-detection DLCZ rate $r_e$ quickly surpasses that of the single-photon protocol for higher detection quantum efficiencies, a field of active technology development~\cite{Hadfield2009}.


\subsection{Hybrid protocol}
\label{sub:expHyb}

\begin{figure}
	\centering
		\includegraphics[width=.45\textwidth]{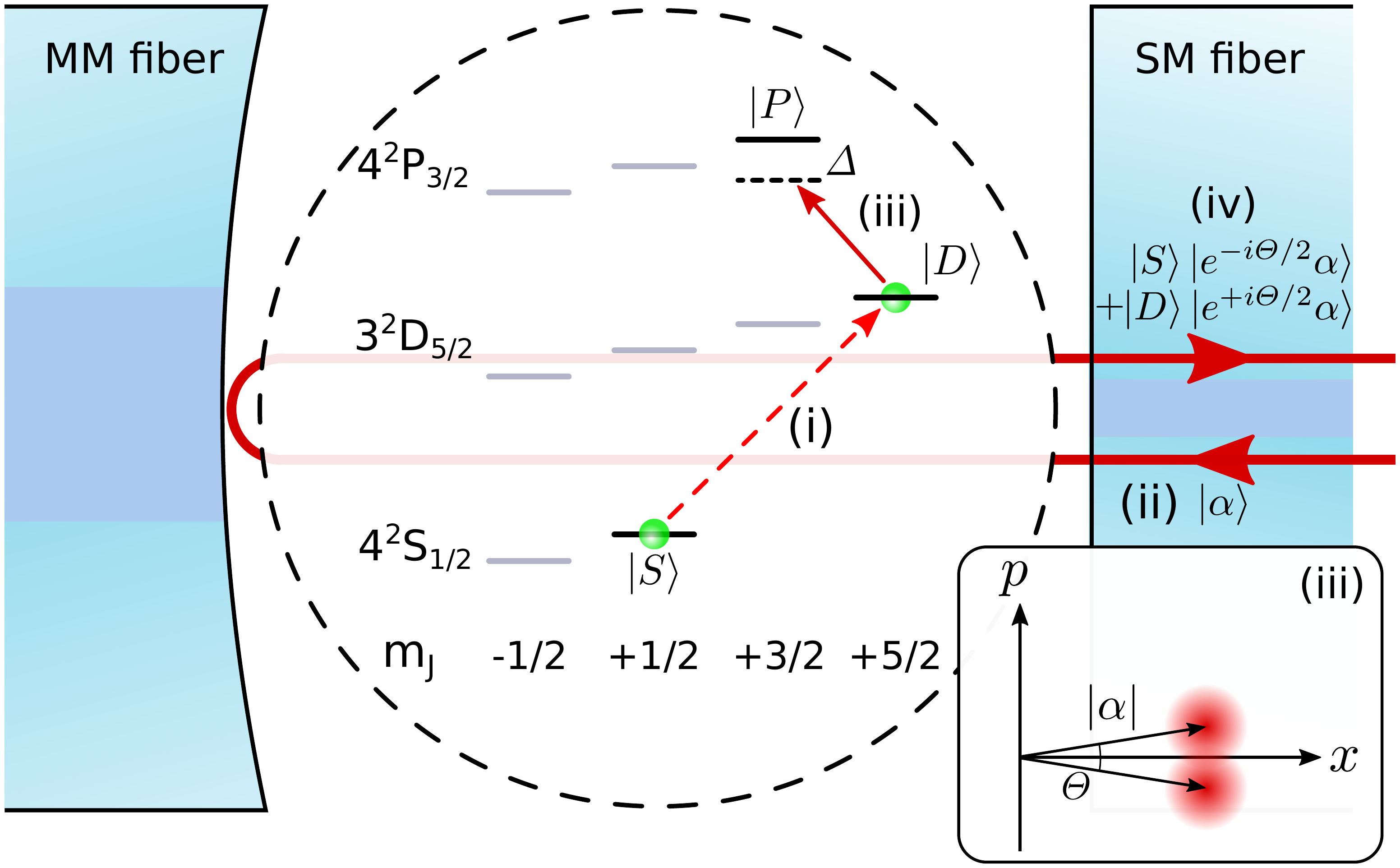}
	\caption{Sketch of the continuous variables hybrid protocol.
(i) The ion is initialized in a superposition of $ \ket{S} $ and $ \ket{D} $, and then shuttled into the cavity mode. Now, the qubus is coupled into the cavity (ii), where both qubus and cavity are detuned by $ \Delta $ to the $ \ket{P}\leftrightarrow\ket{D} $ transition.
The relevant levels of the \Ca-qubit are shown.
(iii) The qubus obtains a phase shift $ \theta $ (see inset) depending on the qubit state, entangling the flying qubus with the stationary qubit in the QR node (iv).}
	\label{fig:hqclvl}       
\end{figure}

An interesting alternative to the previous two protocols, the hybrid protocol as introduced in Sec. \ref{sub:Hyb} employs a continuous variable qubus.
Although the protocol requires an overcoupled cavity, in contrast to the undercoupled one available in our setup, this section will show that upon realization of this condition, the protocol is by far the fastest means of entanglement distribution for medium high finesse.
The qubus is encoded in the phase of a coherent light pulse near $854$~nm, to distribute entanglement between QRs, and the optical qubit of the \Ca-ion (Sec. \ref{sub:qubit}) as stationary qubit in the QR nodes.
The latter is initialized into the superposition (Fig. \ref{fig:hqclvl})

\begin{equation}
	\ket{\Psi_i} = \state{S}{1/2}{+1/2} + \state{D}{5/2}{+5/2} = \ket{S} + \ket{D}
\end{equation}
The ion is then shuttled into the cavity for interaction with the qubus.
The qubus and the local oscillator (LO) reference pulse (Sec. \ref{sub:Hyb}) are created by suitable attenuation and outcoupling of a laser pulse of duration $\tau_q$.
For the generation of entanglement the qubus mode successively interacts dispersively and cavity-enhanced with stationary qubits in distant nodes QR (N) and QR (N+1), see Fig. \ref{fig:theo}c).
The qubus is resonant with the cavity, far detuned by the frequency $\Delta$ from the transition frequency of $\ket{D}\rightarrow\ket{P}$.
The detuning must fulfill $\Delta \gg 2 \cdot g_c =  2 \pi \times 40.2$~MHz, in order to realize the dispersive regime.

Unlike the two previous protocols, the implementation of the hybrid protocol suggests the emplyment of an asymmetric cavity, where the mirror on the I/O fiber features a larger transmittance $T_1=200$~ppm, with $T_i$ ($R_i$) the intensity transmittance (reflectance) of mirror $i$.
All other cavity parameters are assumed to be the same as in Sec.~\ref{sub:cavity}.
The resulting minor decrease of the cooperativity from $C_c \approx 16$ to $C_c \approx 13$ is of no further relevance in this case.

In order to be able to send the coherent pulse from QR~(N) to QR~(N+1), we require an optical circulator.
This can be realized, e.g., by inserting a $\frac{\lambda}{4}$-waveplate between the PBS and the QR nodes in Fig. \ref{fig:theo} c.
Our imperfect cavity incoupling efficiency $\eta_\text{dip}$ reduces the quality of our reflection signal.
However, this detrimental effect can be eliminated by utilizing pulses split off of the LO for error correction, similar to the correction called tuning displacement in \cite{Ladd2006}. 
An LO pulse of correctly chosen amplitude interferes with the qubus at a weak beamsplitter, set after the output port of the optical circulator.
The beamsplitter is set up such that the LO pulse is phase shifted by \textpi, and the unwanted, directly reflected field is subtracted from the qubus.
For a weak beamsplitter, the entangled pulse is almost undisturbed (see \cite{Ladd2006}).
The same error correction must be done after the pulse leaves QR (N+1), before the detection.
For an incoming field $E_\text{inc}$, the field $E_\text{ref}$ reflected from the resonant cavity has the form \cite{meschede_book2005}
\begin{equation}
	E_\text{ref} = \left(\sqrt{R_1} - \sqrt{\varepsilon} \frac{\sqrt{T_1^2 R_2}}{1-\sqrt{R_1 R_2}}\right) E_\text{inc}\,.
\end{equation}
The corrective pulse in this scheme is chosen to be by $E_\text{corr} = \sqrt{R_1} E_\text{inc}\, e^{i \pi}$\,.
Combining both pulses results in the following effective reflection efficiency of the coherent pulse, including mode matching:
\begin{equation}
	\eta_\text{hyb} =\frac{(E_\text{ref} + E_\text{corr})^2}{E_\text{inc}^2} =  \varepsilon \frac{T_1^2 R_2}{(1-\sqrt{R_1 R_2})^2} = 0.25
\end{equation}
for the parameters given above, and the mode matching $\varepsilon = 0.445$.

The qubus pulse duration $\tau_q$ should be suitably long, so that the cavity does not distort the shape of the pulse.
We assume a pulse length of $\tau_q=500$~ns~$\gg \tau_\kappa$, which satisfies this condition.
Due to the length of the qubus, the detection needs to wait the same amount of time, so the entanglement time of the protocol is $\tau_e = 2 \tau_\kappa + \tau_q + \tau_\text{det}\approx 1$~\textmu s, plus the time of light travel between the nodes, with $\tau_\kappa$ the cavity decay time, and $\tau_\text{det}$ the time needed to evaluate the detection events.
As we aim for comparability between the protocols, we set the travel time to zero here.

The total transmission efficiency is 
\begin{equation}
	\eta = \eta_\text{hyb}^2 \, \eta_{FL} \approx 0.06 \, .
\end{equation} 
The effect of absorption due to fiber length, with a transmission efficiency of $\eta_{FL}$, further reduces $\eta$, but is beyond the scope of this paper, and is set to 1 for the remainder of this section.

The protocol is concluded by a homodyne detection, in order to project the joint state of the qubits in both QR nodes onto the mixed Bell state $\mu \ket{\psi^+} + \sqrt{1-\mu^2} \ket{\psi^-}$ of Eqs. \ref{eq:hyb:state} and \ref{eq:hyb:phip}.  
This homodyne detection has to distinguish the phase, see Eq. \ref{eq:Fhomo}. 
Using the estimated parameters from our setup, we find that the fidelity of Bell states is below 0.5 for any value of the distinguishability, and thus for any combination of $\alpha$ and $\theta$, a result of the relatively large transmission loss $\eta$. 

The resulting state is a mixture of the four Bell states, which we cannot distinguish, rendering the homodyne measurement useless for entanglement distribution.

However, we can completely eliminate the \emph{bit-flip} errors that stem from the badly distinguishable phase-rotated and non-phase-rotated parts of Eq. \ref{eq:hyb:state}, by changing the detection scheme to an unambiguous state discrimination setup, as detailed in the following.

\textbf{Unambiguous state discrimination (USD):}
USD \cite{Dusek2000,Loock2008} is an alternative to homodyne detection, where the measurement is set up so that the possible results are \{\emph{definitely entangled, definitely unentangled, unknown}\}, ruling out the possibility of bit-flip errors.
The scheme introduced here is based on Ref. \cite{Loock2008}, and is derived in detail there.

\begin{figure}
	\centering
		\includegraphics[width=.4\textwidth]{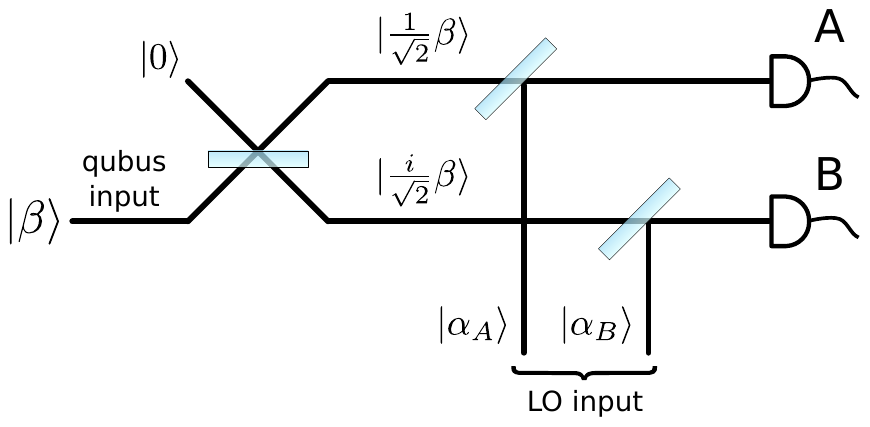}
	\caption{Unambiguous state discrimination setup.
The incoming qubus $\ket{\beta}$, consisting of the superposition of the three coherent states $\ket{\alpha}$, $\ket{\alpha e^{i \theta}}$, and $\ket{\alpha e^{-i \theta}}$, is transformed by a 50/50 beamsplitter into the two output pulses $\ket{\frac{1}{\sqrt{2}}\beta,\frac{i}{\sqrt{2}}\beta}$.
For each output pulse, at another 50/50 beamsplitter a coherent pulse $\ket{\alpha_{\{A/B\}}}$ outcoupled from the LO creates a displacement in phase space. 
}
	\label{fig:usdDet}       
\end{figure}

We are interested in unambiguously identifying the non-phase-shifted part of Eq. \ref{eq:hyb:phip}, which projects the qu\-bits onto the mixed Bell state $\mu \ket{\psi^+} + \sqrt{1-\mu^2} \ket{\psi^-}$.
Fig. \ref{fig:usdDet} shows the detection setup, with an input port for the qubus (the remaining input port is depicted with the vacuum mode), and two output ports to detectors A and B.

The first beamsplitter changes an incoming coherent state $\ket{\beta}$ to
\begin{align}
\label{eq:hybusd}
	\ket{\beta,0}\rightarrow\ket{\frac{1}{\sqrt{2}}\beta,\frac{i}{\sqrt{2}}\beta}\,.
\end{align}

The other two beamsplitters are used to displace the two resulting coherent pulses in phase space by sending phase-shifted coherent pulses $\ket{\alpha_A}$ and $\ket{\alpha_B}$, outcoupled from the LO, into their respective input ports:

\begin{align}
	\hat{D}_A(\alpha_A) &\otimes \hat{D}_B(\alpha_B) \nonumber \\
	 = \hat{D}_A(-\frac{1}{\sqrt{2}}\sqrt{\eta}\alpha e^{i\theta}) &\otimes \hat{D}_B(-\frac{1}{\sqrt{2}}\sqrt{\eta}\alpha e^{-i\theta})\,.
\end{align}

\begin{figure}
	\centering
		\includegraphics[width=.45\textwidth]{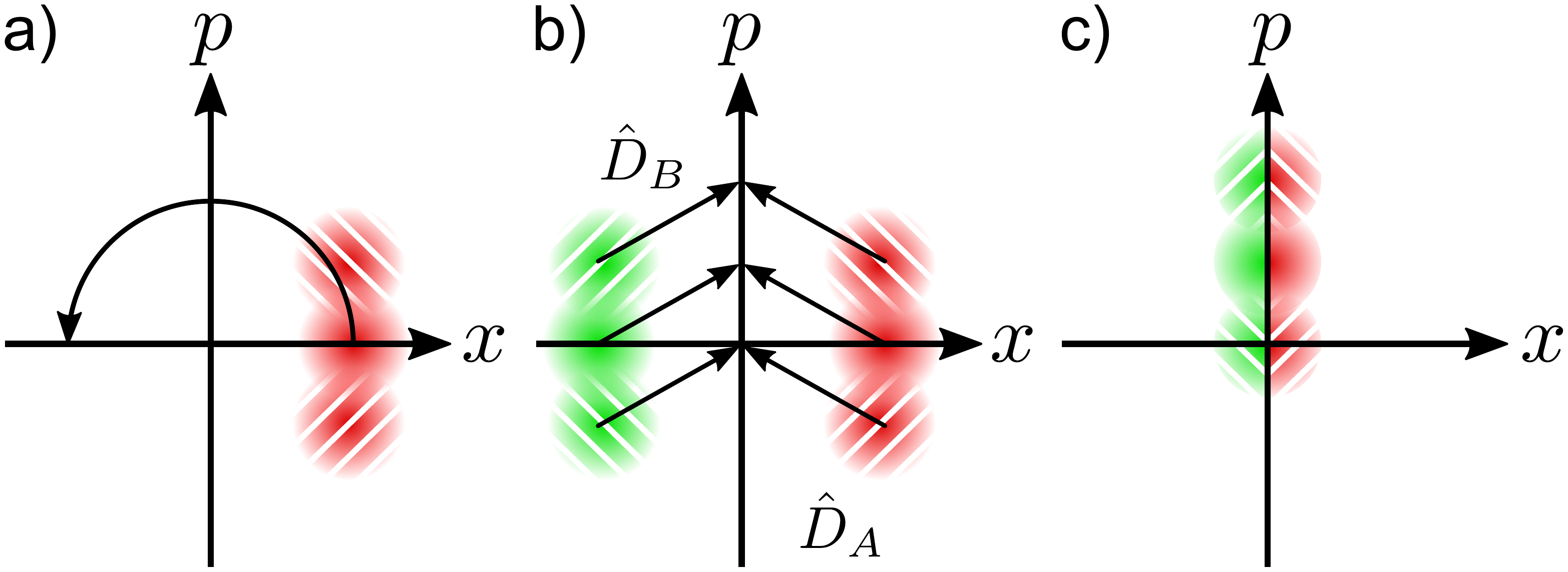}
	\caption{Effect of the USD setup on the qubus in phase space.
a) The qubus, in the incoming port, is in a superposition of $\ket{\alpha}$ (solid), $\ket{\alpha e^{i \theta}}$ (shaded upper left to lower right), and $\ket{\alpha e^{-i \theta}}$ (shaded lower left to upper right). 
b) After the first beamsplitter output port $A$ (red) has states unchanged except for the amplitude (not shown), and output port $B$ (green) has states rotated by $\pi$.
The displacement pulses $\hat{D}_A$ and $\hat{D}_B$ shift the coherent states in phase space such that in each output port, one of the phase shifted states is displaced onto $\ket{0}$.
c) A detection event incompatible with $\ket{0}$ in both detectors A and B is only possible for $\ket{\alpha}$
}
	\label{fig:usdPhaseShift}       
\end{figure}

The displacements are chosen such that for the two phase shifted parts of the wave function $\ket{\sqrt{\eta} \alpha e^{\pm i \theta}}$ as input pulse $\ket{\beta}$, one of the detection ports is always in the vacuum mode (see Fig. \ref{fig:usdPhaseShift}).

This can be seen by applying these transformations to the three different qubus input states from Eq. \ref{eq:hyb:phip}:
\begin{samepage}
\begin{flalign}
\label{eq:usdDet}
	\ket{\sqrt{\eta} \alpha,0} &\rightarrow \ket{\frac{1}{\sqrt{2}} \sqrt{\eta} \alpha(1-e^{i\theta}), \frac{1}{\sqrt{2}} \sqrt{\eta} \alpha(1-e^{-i\theta})}\,, \nonumber \\
	\ket{\sqrt{\eta} \alpha e^{i\theta},0} &\rightarrow	\ket{0,\frac{1}{\sqrt{2}} \sqrt{\eta} \alpha 2 i \sin{\theta}} \,,\\
	\ket{\sqrt{\eta} \alpha e^{-i\theta},0} &\rightarrow	\ket{-\frac{1}{\sqrt{2}} \sqrt{\eta} \alpha 2 i \sin{\theta},0}\,. \nonumber
\end{flalign}
\end{samepage}

Of the 4 possible detector click patterns, only both detectors firing in coincidence definitely identifies the entangled part of Eq. \ref{eq:hyb:state}.

The success probability for this event is given by
\begin{align}
\label{eq:PUSD}
	P_\text{e} = \frac{1}{2}\big( 1 - e^{-\eta\alpha^2 (1-\cos{\theta})} \big)^2
\end{align}
with a fidelity of the final state
\begin{align}
\label{eq:FUSD}
	F=\bra{\psi^+}\rho\ket{\psi^+}=\mu^2\,.
\end{align}

Note that $P_e$ already includes the effect of the USD setup, e.g. the signal reduction by the first beamsplitter of Fig. \ref{fig:usdDet}.
It was assumed that the signal remains strong enough to be clearly discerned from the vacuum state, which should be easily attainable with modern detectors and a moderate pulse amplitude $\alpha$.
The final repetition rate of entanglement distribution is given by
\begin{equation}
\label{eq:USDrate}
	r_e = \frac{P_e}{\tau_\text{run}}\,,
\end{equation}
for a time per experimental run $\tau_\text{run}$.

\textbf{Entanglement generation rates:}
Similar to the other two protocols, one has to find a trade off between fidelity and efficiency. 
Choosing a fidelity $F=0.99$ for good comparison with the DLCZ schemes, Eq. \ref{eq:PUSD} gives us a success probability of $P_e \approx 9 \times 10^{-7}$ for optimal values of the distiguishability $d$, Eq. \ref{eq:hyb:d}.
In order to find the optimal parameters, we require that $F=\mu^2=.99$ (Eq. \ref{eq:hybF}), and maximize $P_e$ with this constraint, with respect to $\alpha$ and $\theta$.
For, e.g., $\alpha=100$, the optimal phase shift angle is $\theta \approx 2 \times 10^{-3}$.
Thus, there are two ways to experimentally achieve the optimal rate:
We can either change the qubus intensity $|\alpha|^2$ to optimize $P_e$ for a given interaction strength between qubus and qubit, or we can move the ion within the cavity field, and thus change $\theta$, to optimize $P_e$ for a given qubus intensity $\alpha$.

Since the dispersive interaction does not disturb the ionic state, we can skip ion initializations in between tries until a new cooling cycle is necessary, greatly reducing the time investment required per run as compared to the other two protocols, setting $\tau_\text{prep}=0$.
These assumptions lead to a mean time to entanglement of $\tau_e \approx 0.13$~s, and a  entanglement generation rate of $r \approx 8$~s$^{-1}$, which includes the ion initialization.
Already for a modest drop in fidelity to 0.95, the rate increases to 170~s$^{-1}$, while for $F=0.8$ the rate is $\approx 700$~s$^{-1}$.
The initialization and state readout errors once again reduce the fidelities presented in this section by a factor of $0.995$\,.

\subsection{Performance comparison of the protocols}

\begin{figure}
	\centering
		\includegraphics[width=.48\textwidth]{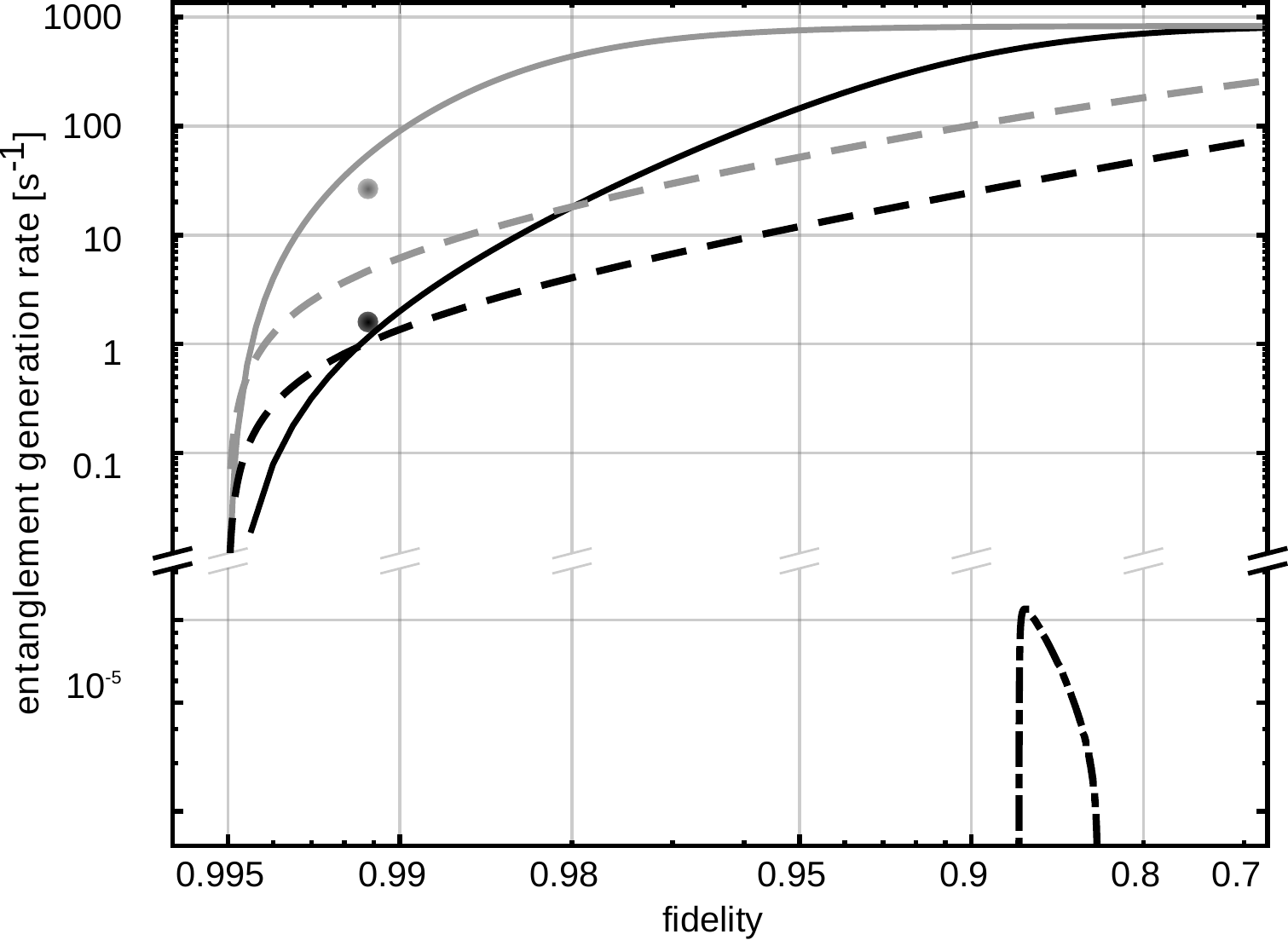}
	\caption{Entanglement generation rates possible at a given fidelity for each protocol. The EPR protocol (dash-dotted), the single-photon DLCZ protocol (dashed), the two-photon DLCZ protocol (dot), use the undercoupled cavity. The hybrid protocol with USD, for an overcoupled but otherwise identical cavity, is drawn solid. For high fidelities, $F \gtrsim 0.99$, the DLCZ protocol shows the best performance for our apparatus, while at fidelities below 0.99 the hybrid protocol becomes visibly better. For higher quantum efficiencies and an optimized cavity (see text), upper bounds for hybrid and DLCZ protocol rates are also plotted (grey). Results have been obtained numerically from Eqs. \ref{eq:eprre}, \ref{eq:dlczrate} and \ref{eq:USDrate}}
	\label{fig:FvR}       
\end{figure}

In the following we discuss key results for all protocols.
The entanglement generation rate as a function of the fidelity is plotted for all considered protocol versions in Fig. \ref{fig:FvR}.
Typically, there is an experimental parameter that can be tuned to trade the fidelity of resulting Bell states for their production rate.
Also, the protocols differ considerably in their requirements for inter-node phase coherence of lasers used for state manipulation of the ion, and  for the stability of the connecting fiber link.

The tunable parameter for the \emph{EPR protocol} is the detection window $\tau_W$, i.e. the time to wait for re-initiali\-zation. 
A short $\tau_W$ up to a point increases the fidelity by making erroneous dark counts less likely (see Eqs. \ref{eq:2dc} and \ref{eq:hdc}), while reducing the success probability by stopping the protocol before any EPR photon has interacted with an ion, see Eq. \ref{eq:EPRRate}\,.
Choosing this parameter too short, however, reduces the rate quickly as the waiting time becomes shorter than the temporal shape of the herald photon wavepackets leaving the cavities.
The interferometric stability of optical frequencies between QR nodes is not required in this protocol: 
The phase imprinted on the initial ionic state is set by the relative phase of the two transitions near 729~nm at each node (see Fig. \ref{fig:eprlvl}), whose frequency difference lies in the RF range. 
Consequently, phase coherence can be attained by distributing a stable RF reference signal between neighboring nodes. 
No interferometric stability of the fiber link required, as any phase collected by the entangled photonic state during transmission is a global one. 
In order to improve the EPR protocol, one would work on the two major inefficiencies: 
On the one hand, the small photon detection probability of the herald is the largest contributor to the slow entanglement generation rate. 
Recent work (e.g., ~\cite{Kurz2014,Schug2014}) uses high aperture laser objectives for improved herald collection efficiency. 
The detection could also be improved by using a high-finesse, dual wavelength cavity for both the detection wavelength near $393$~nm light, and the EPR-pair wavelength near 854~nm. 
Enhanced emission through the Purcell effect, however, would require a strong coupling regime for UV-cavities and come at the price of technical complexity. 

The \emph{single-photon detection DLCZ} offers as tunable parameter the single photon generation rate, see Eqs. \ref{eq:dlcz1suc} and \ref{eq:dlcz:1suc}.  The rate of Bell pair production rises with a higher rate of single photons, while the fidelity drops caused from a higher probability for simultaneous photon emission in both QR nodes.
The absolute phase of the laser field creating the qubit (see Fig. \ref{fig:dlczlvl1} ) cannot be controlled, such that phase coherence between neighboring nodes needs to be established at optical frequencies. The relative phase of the photonic state depends on the position of the detector~\cite{Cabrillo1999} which requires that the fiber link between QR nodes must be interferometrically stabilized. For details concerning the interferometric stability of fiber links, see e.g.~\cite{Sangouard2011,Ma1994,Minar2008,Jiang2008}\,.
The rate of entangled pairs would benefit mostly from an improved transfer of the entangled photon to the detection apparatus, which is typically mainly limited by the coupling efficiency between the fiber cavity mode and the fiber mode. We realistically aim for increasing the mode matching $\varepsilon$ from $0.44$ to $\approx 0.54$\,. 
Ultimately we are limited by the geometry of our setup to $\varepsilon_\text{max} \approx 0.57$\,. 
For small values of the transfer and detection probability $P_\text{det}$, the entanglement rates increase only linearly with mode matching, by at most $\approx 23$~\%\,. Similar gains could be achieved by reducing mirror losses to optimize the impedance matching, Eq. \ref{eq_eta-impedance}, or by optimizing the ratio of reflectivity of the two cavity mirrors to improve the cavity outcoupling coefficient $\eta_\text{out}$.

For the \emph{two-photon DLCZ}, decreasing the rate does not improve the  fidelity of Bell states. Consequently, its performance is depicted in Fig. \ref{fig:FvR} by single points.
In this protocol, the ionic state is created by a bichromatic optical field (see Fig. 9), which allows for a phase control by an RF-link. 
During the photon transmission only the global phase of the photonic state is altered, which means the stability of the fiber link needs to ensure temporal coincidence of both of the photons impinging at the beamsplitter, but interferometric stability is not required~\cite{Sangouard2011}.
Any improvement of the mode matching would be most important for this protocol, as this parameter enters quadratically in Eq. \ref{eq:dlcz2suc}. 
A gain in performance (up to a 51~\% increase for the optimized mode matching) is expected for a more precisely aligned fiber cavity.
%

Concerning the \emph{hybrid protocol}, the larger the parameters $\theta$ or $\alpha$ are chosen, the more the rate is increased, at the cost of fidelity. 
This follows from Eqs. \ref{eq:hybF} and \ref{eq:PUSD} which mirrors the mixing of pure and entangled states described in Sec.~\ref{sub:Hyb}.
The protocol needs optically phase-stable lasers, as a monochromatic field is used to create the initial ionic state (see Fig. \ref{fig:hqclvl}). 
Interferometric stability for the fiber link is not required: 
The quantum information in the qubus is transmitted together with a local oscillator pulse, and a homodyne-type measurement eliminates all phases collected by both those parts during transmission, also in the case of USD detection.
The hybrid protocol would benefit, similar to the DLCZ protocols, from an improved the mode matching between fiber cavity and fiber, as well as an improved impedance matching. 
Realistically, we could improve the mode matching to $\varepsilon=0.54$ which would yield an almost two orders of magnitude larger entanglement generation rate with $F = 0.99$. 

In the hybrid and single-photon DLCZ protocols we might alleviate the requirement for inter-node phase coherence of lasers with an alternative way of creating the superposition of the $\ket{S}$- and $\ket{D}$-states: Creating a coherent superposition between the two $S_{1/2}$-Zeeman states using a Raman laser interaction (see Fig. \ref{fig:lvl})~\cite{Poschinger2009}, followed by a coherent population transfer from one of those sublevels to the D-state via rapid adiabatic passage (RAP). 
After entanglement as been generated, each qubit could be coherently returned to the spin qubit by a second RAP, before the necessary local operations are performed. 
The phase coherence between QR nodes would then be ensured by a RF reference exchange.

In conclusion and taking above discussion into account, we note that the EPR protocol shows a comparatively small rate of entanglement generation and can hardly be implemented in our ion trap-cavity platform. However, for both the DLCZ protocol and the hybrid protocol utilizing USD, the estimated rates and fidelities for our parameters suggest the possibility of outperforming the state-of-the-art \cite{Hucul2014}, where free-space photon collection is employed. Both protocols would profit from a fiber optical cavity with a high reflectance end mirror on the multimode fiber~\cite{Steiner2014}. If the losses in the mirror coatings and the transmission through the high reflectance mirror could be brought well below 50~ppm, the cavity outcoupling coefficient would improve to a value that we estimate with $\eta_\text{out} \approx 0.26$. This would double $\eta_\text{out}$ and in turn $P_\text{det}$ for both DLCZ protocols, see Eq. \ref{eq:dlczdet}. Furthermore, with $\varepsilon \approx 0.54$, these conditions could let the hybrid USD protocol reach a rate of $\approx 750$~s$^{-1}$ for a state fidelity of $F=0.95$ (Fig. \ref{fig:FvR}).

\section{Conclusion and outlook}
\label{outlook}

We discussed the implementations of three quantum repeater protocols, namely the distributed EPR protocol, the DLCZ protocol and the hybrid protocol, and various protocol extensions with an ion trap setup. When comparing the protocols, we find that the DLCZ protocol and the hybrid protocol outperform the EPR protocol. The most important limitations are of technical nature. We see that improving the mode matching from the experimentally determined value of 0.44 to 0.54, which can be achieved through an improved cavity alignement, entanglement generation rates of 30~s$^{-1}$ at fidelities of 0.9 with the DLCZ protocol are within reach. Furthermore, the impedance matching of the fiber optical cavity may be improved by coating the mirror of the I/O fiber with a different transmission coefficient as compared to the high reflection end mirror on the multimode fiber, which would allow a rate of up to 60~s$^{-1}$.

The hybrid protocol appears interesting for future investigations, even though we would need to change the fiber cavity: The quick entanglement generation rate at medium-high fidelity makes it an ideal candidate to be used in combination with entanglement distillation schemes~\cite{Bennett1996,Duan2000,Pan2001}. That requires a rate of entanglement generation exceeding the decay rate of the stationary quantum memory. 
The qubit's coherence time is, in our case, limited by magnetic field fluctuations to 10~ms, which means that for an optimized hybrid rate of up to $750$~s$^{-1}$ at $F=0.95$, already more entanglement per time would be created than lost. Encoding the stationary qubit in a decoherence-free substate~\cite{Haeffner2005} and executing the mapping between qubus and logical qubit comprised of a two-ion Bell state~\cite{Casabone2015} with a coherence time on the order of 10~s would improve the ratio of entanglement distribution time to coherence time to $\approx 7500$\,.

Yet another option is the use of error correction codes in entanglement distribution. The basic code in \cite{Calderbank1996,Schindler2011} requires 2 ancilla qubits per qubit, few enough to be realized in our setup, without the need to build multiple traps per QR node.

Furthermore, combining our platform with a single photon wavelength converter to provide compatibility with telecom fibers opens up a promising perspective of evolving our system from a lab based proof-of-principle experiment to a prototype, which may increase the maximally achievable distance for quantum communication. 
We thank Peter van Loock, Denis Gonta and Pascal Eich for helpful discussions.
We acknowledge financial support by the European commission within the IP SIQS and by the Bundesministerium f\"ur Bildung und For\-schung via IKT 2020 (Q.com).

%
%

%


\begin{thebibliography}{10}

\bibitem{Duan2001}
L.-M. Duan, M.~D. Lukin, J.~I. Cirac, and P.~Zoller.
\newblock Long-distance quantum communication and with atomic ensembles and
  linear optics.
\newblock {\em Nature}, 414:413--418, Nov 2001.

\bibitem{Loock2006}
P.~van Loock, T.~D. Ladd, K.~Sanaka, F.~Yamaguchi, K.~Nemoto, W.~J. Munro, and
  Y.~Yamamoto.
\newblock Hybrid quantum repeater using bright coherent light.
\newblock {\em Phys. Rev. Lett.}, 96:240501, Jun 2006.

\bibitem{1367-2630-15-8-085004}
N.~Sangouard, J.-D. Bancal, P.~M\"uller, J.~Ghosh, and J.~Eschner.
\newblock Heralded mapping of photonic entanglement into single atoms in free
  space: proposal for a loophole-free {B}ell test.
\newblock {\em New J. Phys.}, 15(8):085004, 2013.

\bibitem{Wooters1982}
W.~K. Wooters and W.~H. Zurek.
\newblock A single quantum cannot be cloned.
\newblock {\em Nature}, 229:802--803, Oct 1982.

\bibitem{Bennett1984}
C.~H. Bennett and G.~Brassard.
\newblock Quantum cryptography: Public key distribution and coin tossing.
\newblock {\em Proceedings of the IEEE International Conference on Computers,
  Systems, and Signal Processing}, page 175, 1984.

\bibitem{Ekert1991}
A.~K. Ekert.
\newblock Quantum cryptography based on {B}ell's theorem.
\newblock {\em Phys. Rev. Lett.}, 67:661--663, Aug 1991.

\bibitem{IDQuantique}
{ID Quantique}.
\newblock http://www.idquantique.com/.

\bibitem{magiQ}
{MagiQ Technologies}.
\newblock http://www.magiqtech.com/.

\bibitem{Ma2012a}
X.-S. Ma, T.~Herbst, T.~Scheidl, D.~Wang, S.~Kropatschek, W.~Naylor,
  B.~Wittmann, A.~Mech, J.~Kofler, E.~Anisimova, V.~Makarov, T.~Jennewein,
  R.~Ursin, and A.~Zeilinger.
\newblock Quantum teleportation over 143 kilometres using active feed-forward.
\newblock {\em Nature}, 489(7415):269--273, Sep 2012.

\bibitem{Vallone2015}
G.~Vallone, D.~Bacco, D.~Dequal, S.~Gaiarin, V.~Luceri, G.~Bianco, and
  P.~Villoresi.
\newblock Experimental satellite quantum communications.
\newblock {\em Phys. Rev. Lett.}, 115:040502, Jul 2015.

\bibitem{Briegel1998}
H.-J. Briegel, W.~D\"ur, J.~I. Cirac, and P.~Zoller.
\newblock Quantum repeaters: The role of imperfect local operations in quantum
  communication.
\newblock {\em Phys. Rev. Lett.}, 81:5932--5935, Dec 1998.

\bibitem{Bennett1996}
C.~H. Bennett, G.~Brassard, S.~Popescu, B.~Schumacher, J.~A. Smolin, and W.~K.
  Wootters.
\newblock Purification of noisy entanglement and faithful teleportation via
  noisy channels.
\newblock {\em Phys. Rev. Lett.}, 76:722--725, Jan 1996.

\bibitem{Calderbank1996}
A.~R. Calderbank and P.~W. Shor.
\newblock Good quantum error-correcting codes exist.
\newblock {\em Phys. Rev. A}, 54(2):1098--1105, Aug 1996.

\bibitem{Reichle2006}
R.~Reichle, D.~Leibfried, E.~Knill, J.~Britton, R.~B. Blakestad, J.~D. Jost,
  C.~Langer, R.~Ozeri, S.~Seidelin, and D.~J. Wineland.
\newblock Experimental purification of two-atom entanglement.
\newblock {\em Nature}, 443(7113):838--841, October 2006.

\bibitem{Terhal2015}
B.~M. Terhal.
\newblock Quantum error correction for quantum memories.
\newblock {\em Rev. Mod. Phys.}, 87(2):307--346, Apr 2015.

\bibitem{Munro2012}
W.~J. Munro, A.~M. Stephens, S.~J. Devitt, K.~A. Harrison, and K.~Nemoto.
\newblock Quantum communication without the necessity of quantum memories.
\newblock {\em Nat. Photonics}, 6(11):777--781, Oct 2012.

\bibitem{Pirandola2015}
S.~Pirandola, J.~Eisert, C.~Weedbrook, A.~Furusawa, and S.~L. Braunstein.
\newblock Advances in quantum teleportation.
\newblock {\em ArXiv e-prints}, arxiv:1505.07831, May 2015.

\bibitem{Kuzmich2003}
A.~Kuzmich, W.~P. Bowen, A.~D. Boozer, A.~Boca, C.~W. Chou, L.-M. Duan, and
  H.~J. Kimble.
\newblock Generation of nonclassical photon pairs for scalable quantum
  communication with atomic ensembles.
\newblock {\em Nature}, 423(6941):731--734, Jun 2003.

\bibitem{Duan2010}
L.-M. Duan and C.~Monroe.
\newblock Colloquium: Quantum networks with trapped ions.
\newblock {\em Rev. Mod. Phys.}, 82(2):1209--1224, Apr 2010.

\bibitem{Ritter2012}
S.~Ritter, C.~Nolleke, C.~Hahn, A.~Reiserer, A.~Neuzner, M.~Uphoff, M.~Mucke,
  E.~Figueroa, J.~Bochmann, and G.~Rempe.
\newblock An elementary quantum network of single atoms in optical cavities.
\newblock {\em Nature}, 484(7393):195--200, April 2012.

\bibitem{Hucul2014}
D.~Hucul, I.~V. Inlek, G.~Vittorini, C.~Crocker, S.~Debnath, S.~M. Clark, and
  C.~Monroe.
\newblock Modular entanglement of atomic qubits using photons and phonons.
\newblock {\em Nat. Phys.}, 11(1):37--42, Nov 2014.

\bibitem{Press2008}
D.~Press, T.~D. Ladd, B.~Zhang, and Y.~Yamamoto.
\newblock Complete quantum control of a single quantum dot spin using ultrafast
  optical pulses.
\newblock {\em Nature}, 456(7219):218--221, Nov 2008.

\bibitem{Gschrey2015}
M.~Gschrey, A.~Thoma, P.~Schnauber, M.~Seifried, R.~Schmidt, B.~Wohlfeil,
  L.~Kruger, J.-H. Schulze, T.~Heindel, S.~Burger, F.~Schmidt, A.~Strittmatter,
  S.~Rodt, and S.~Reitzenstein.
\newblock Highly indistinguishable photons from deterministic quantum-dot
  microlenses utilizing three-dimensional in situ electron-beam lithography.
\newblock {\em Nat. Commun.}, 6, Jul 2015.

\bibitem{Mundt2002}
A.~B. Mundt, A.~Kreuter, C.~Becher, D.~Leibfried, J.~Eschner, F.~Schmidt-Kaler,
  and R.~Blatt.
\newblock Coupling a single atomic quantum bit to a high finesse optical
  cavity.
\newblock {\em Phys. Rev. Lett.}, 89:103001, Aug 2002.

\bibitem{Steiner2014}
M.~Steiner, H.~M. Meyer, J.~Reichel, and M.~K\"ohl.
\newblock Photon emission and absorption of a single ion coupled to an
  optical-fiber cavity.
\newblock {\em Phys. Rev. Lett.}, 113(26), Dec 2014.

\bibitem{Casabone2015}
B.~Casabone, K.~Friebe, B.~Brandst\"atter, K.~Sch\"uppert, R.~Blatt, and
  E.~Northup, T.\.
\newblock Enhanced quantum interface with collective ion-cavity coupling.
\newblock {\em Phys. Rev. Lett.}, 114:023602, Jan 2015.

\bibitem{Ikuta_2011}
R.~Ikuta, Y.~Kusaka, T.~Kitano, H.~Kato, T.~Yamamoto, M.~Koashi, and N.~Imoto.
\newblock Wide-band quantum interface for visible-to-telecommunication
  wavelength conversion.
\newblock {\em Nat Comms}, 2:1544, Nov 2011.

\bibitem{Zaske2012}
S.~Zaske, A.~Lenhard, C.~A. Ke\ss{}ler, J.~Kettler, C.~Hepp, C.~Arend,
  R.~Albrecht, W.-M. Schulz, M.~Jetter, P.~Michler, and C.~Becher.
\newblock Visible-to-telecom quantum frequency conversion of light from a
  single quantum emitter.
\newblock {\em Phys. Rev. Lett.}, 109:147404, Oct 2012.

\bibitem{Brassard1994}
G.~Brassard and L.~Salvail.
\newblock Secret-key reconciliation by public discussion.
\newblock In {\em Advances in Cryptology {\textemdash} {EUROCRYPT} '93}, pages
  410--423. Springer Science \& Business Media, 1994.

\bibitem{Bennett1988}
C.~H. Bennett, G.~Brassard, and J.-M. Robert.
\newblock Privacy amplification by public discussion.
\newblock {\em SIAM J. Comput.}, 17(2):210--229, Apr 1988.

\bibitem{Deutsch1996}
D.~Deutsch, A.~Ekert, R.~Jozsa, C.~Macchiavello, S.~Popescu, and A.~Sanpera.
\newblock Quantum privacy amplification and the security of quantum
  cryptography over noisy channels.
\newblock {\em Phys. Rev. Lett.}, 77:2818--2821, Sep 1996.

\bibitem{Lloyd2001}
S.~Lloyd, M.~S. Shahriar, J.~H. Shapiro, and P.~R. Hemmer.
\newblock Long distance, unconditional teleportation of atomic states via
  complete {B}ell state measurements.
\newblock {\em Phys. Rev. Lett.}, 87:167903, Sep 2001.

\bibitem{Cabrillo1999}
C.~Cabrillo, J.~I. Cirac, P.~Garcia-Fernandez, and P.~Zoller.
\newblock Creation of entangled states of distant atoms by interference.
\newblock {\em Phys. Rev. A}, 59:1025--1033, Feb 1999.

\bibitem{Simon2003}
C.~Simon and W.~T.~M. Irvine.
\newblock Robust long-distance entanglement and a loophole-free {B}ell test
  with ions and photons.
\newblock {\em Phys. Rev. Lett.}, 91:110405, Sep 2003.

\bibitem{Kurz2014}
C.~Kurz, M.~Schug, P.~Eich, J.~Huwer, P.~M\"uller, and J.~Eschner.
\newblock Experimental protocol for high-fidelity heralded photon-to-atom
  quantum state transfer.
\newblock {\em Nat. Commun.}, 5:--, November 2014.

\bibitem{Schug2014}
M.~Schug, C.~Kurz, P.~Eich, J.~Huwer, P.~M\"uller, and J.~Eschner.
\newblock Quantum interference in the absorption and emission of single photons
  by a single ion.
\newblock {\em Phys. Rev. A}, 90:023829, Aug 2014.

\bibitem{Stute2012}
A.~Stute, B.~Casabone, P.~Schindler, T.~Monz, P.~O. Schmidt, B.~Brandst\"atter,
  T.~E. Northup, and R.~Blatt.
\newblock Tunable ion-photon entanglement in an optical cavity.
\newblock {\em Nature}, 485(7399):482--485, May 2012.

\bibitem{Zippilli2008}
S.~Zippilli, G.~A. Olivares-Rentería, G.~Morigi, C.~Schuck, F.~Rohde, and
  J.~Eschner.
\newblock Entanglement of distant atoms by projective measurement: the role of
  detection efficiency.
\newblock {\em New J. Phys.}, 10(10):103003, 2008.

\bibitem{Ladd2006}
T.~D. Ladd, P.~van Loock, K.~Nemoto, W.~J. Munro, and Y.~Yamamoto.
\newblock Hybrid quantum repeater based on dispersive cqed interactions between
  matter qubits and bright coherent light.
\newblock {\em New J. Phys.}, 8(9):184, 2006.

\bibitem{Loock2008}
P.~van Loock, N.~L\"utkenhaus, W.~J. Munro, and K.~Nemoto.
\newblock Quantum repeaters using coherent-state communication.
\newblock {\em Phys. Rev. A}, 78:062319, Dec 2008.

\bibitem{Bennett1993}
C.~H. Bennett, G.~Brassard, C.~Cr\'epeau, R.~Jozsa, A.~Peres, and W.~K.
  Wootters.
\newblock Teleporting an unknown quantum state via dual classical and
  {Einstein}-podolsky-rosen channels.
\newblock {\em Phys. Rev. Lett.}, 70(13):1895--1899, Mar 1993.

\bibitem{Riebe2008}
M.~Riebe, T.~Monz, K.~Kim, A.~S. Villar, P.~Schindler, M.~Chwalla, M.~Hennrich,
  and R.~Blatt.
\newblock Deterministic entanglement swapping with an ion-trap quantum
  computer.
\newblock {\em Nat. Phys.}, 4:839--842, 11 2008.

\bibitem{Schulz2008}
S.~A. Schulz, U.~Poschinger, F.~Ziesel, and F.~Schmidt-Kaler.
\newblock Sideband cooling and coherent dynamics in a microchip multi-segmented
  ion trap.
\newblock {\em New J. Phys.}, 10(4):045007, 2008.

\bibitem{Poschinger2009}
U.~G. Poschinger, G.~Huber, F.~Ziesel, M.~Deiss, M.~Hettrich, S.~A. Schulz,
  G.~Poulsen, M.~Drewsen, R.~J. Hendricks, K.~Singer, and F.~Schmidt-Kaler.
\newblock Coherent manipulation of a $^{40}${Ca}$^+$ spin qubit in a micro ion
  trap.
\newblock {\em J. Phys. B}, 42:154013, 2009.

\bibitem{Schulz2006}
S.~Schulz, U.~Poschinger, K.~Singer, and F.~Schmidt-Kaler.
\newblock Optimization of segmented linear paul traps and transport of stored
  particles.
\newblock {\em Fortschr. Phys.}, 54(8-10):648--665, 2006.

\bibitem{Hunger2010}
D.~Hunger, T.~Steinmetz, Y.~Colombe, C.~Deutsch, T.~W. H{\"a}nsch, and
  J.~Reichel.
\newblock A fiber fabry-p\'erot cavity with high finesse.
\newblock {\em New J. Phys.}, 12(6):065038, 2010.

\bibitem{Walther2012}
A.~Walther, F.~Ziesel, T.~Ruster, S.~T. Dawkins, K.~Ott, M.~Hettrich,
  K.~Singer, F.~Schmidt-Kaler, and U.~Poschinger.
\newblock Controlling fast transport of cold trapped ions.
\newblock {\em Phys. Rev. Lett.}, 109:080501, Aug 2012.

\bibitem{Bowler2012}
R.~Bowler, J.~Gaebler, Y.~Lin, T.~R. Tan, D.~Hanneke, J.~D. Jost, J.~P. Home,
  D.~Leibfried, and D.~J. Wineland.
\newblock Coherent diabatic ion transport and separation in a multizone trap
  array.
\newblock {\em Phys. Rev. Lett.}, 109:080502, Aug 2012.

\bibitem{Kaufmann2014}
H.~Kaufmann, T.~Ruster, C.~T. Schmiegelow, F.~Schmidt-Kaler, and U.~G.
  Poschinger.
\newblock Dynamics and control of fast ion crystal splitting in segmented paul
  traps.
\newblock {\em New J. Phys.}, 16(7):073012, 2014.

\bibitem{Ruster2014}
T.~Ruster, C.~Warschburger, H.~Kaufmann, C.~T. Schmiegelow, A.~Walther,
  M.~Hettrich, A.~Pfister, V.~Kaushal, F.~Schmidt-Kaler, and U.~G. Poschinger.
\newblock Experimental realization of fast ion separation in segmented paul
  traps.
\newblock {\em Phys. Rev. A}, 90:033410, 2014.

\bibitem{1367-2630-15-12-123012}
P.~Schindler, D.~Nigg, T.~Monz, J.~T. Barreiro, E.~Martinez, S.~X. Wang,
  S.~Quint, M.~F. Brandl, V.~Nebendahl, C.~F. Roos, M.~Chwalla, M.~Hennrich,
  and R.~Blatt.
\newblock A quantum information processor with trapped ions.
\newblock {\em New J. Phys.}, 15(12):123012, 2013.

\bibitem{Leibfried2003a}
D.~Leibfried, B.~DeMarco, V.~Meyer, D.~Lucas, M.~Barrett, J.~Britton, W.~M.
  Itano, B.~Jelenkovic, C.~Langer, T.~Rosenband, and D.~J. Wineland.
\newblock Experimental demonstration of a robust, high-fidelity geometric two
  ion-qubit phase gate.
\newblock {\em Nature}, 422(6930):412--415, March 2003.

\bibitem{Colombe2007}
Y.~Colombe, T.~Steinmetz, G.~Dubois, F.~Linke, D.~Hunger, and J.~Reichel.
\newblock Strong atom-field coupling for bose-einstein condensates in an
  optical cavity on a chip.
\newblock {\em Nature}, 450(7167):272--276, November 2007.

\bibitem{Brandstaetter2013}
B.~{Brandst{\"a}tter}, A.~{McClung}, K.~{Sch{\"u}ppert}, B.~{Casabone},
  K.~{Friebe}, A.~{Stute}, P.~O. {Schmidt}, C.~{Deutsch}, J.~{Reichel},
  R.~{Blatt}, and T.~E. {Northup}.
\newblock Integrated fiber-mirror ion trap for strong ion-cavity coupling.
\newblock {\em Rev. Sci. Instrum.}, 84(12), 2013.

\bibitem{Harlander2010}
M.~Harlander, M.~Brownnutt, W.~H{\"a}nsel, and R.~Blatt.
\newblock Trapped-ion probing of light-induced charging effects on dielectrics.
\newblock {\em New J. Phys.}, 12(9):093035, 2010.

\bibitem{Herskind2011}
P.~F. Herskind, S.~X. Wang, M.~Shi, Y.~Ge, M.~Cetina, and I.~L. Chuang.
\newblock Microfabricated surface ion trap on a high-finesse optical mirror.
\newblock {\em Opt. Lett.}, 36(16):3045, 2011.

\bibitem{Brownnutt2014}
M.~{Brownnutt}, M.~{Kumph}, P.~{Rabl}, and R.~{Blatt}.
\newblock Ion-trap measurements of electric-field noise near surface.
\newblock {\em ArXiv e-prints}, arXiv:1409.6572, September 2014.

\bibitem{Gallego2015}
J.~Gallego, S.~Ghosh, S.~K. Alavi, W.~Alt, M.~Martinez-Dorantes, D.~Meschede,
  and L.~Ratschbacher.
\newblock High finesse fiber fabry-perot cavities: Stabilization and mode
  matching analysis.
\newblock August 2015.

\bibitem{meschede_book2005}
D.~Meschede.
\newblock {\em Optik, Licht und Laser}.
\newblock Lehrbuch Physik. Teubner, 2005.

\bibitem{Hood2001}
C.~J. Hood, H.~J. Kimble, and J.~Ye.
\newblock Characterization of high-finesse mirrors: Loss, phase shifts, and
  mode structure in an optical cavity.
\newblock {\em Phys. Rev. A}, 64:033804, Aug 2001.

\bibitem{Schuck2010}
C.~Schuck, F.~Rohde, N.~Piro, M.~Almendros, J.~Huwer, M.~W. Mitchell,
  M.~Hennrich, A.~Haase, F.~Dubin, and J.~Eschner.
\newblock Resonant interaction of a single atom with single photons from a
  down-conversion source.
\newblock {\em Phys. Rev. A}, 81:011802, Jan 2010.

\bibitem{Huwer2013}
J.~Huwer, J.~Ghosh, N.~Piro, M.~Schug, F.~Dubin, and J.~Eschner.
\newblock Photon entanglement detection by a single atom.
\newblock {\em New J. Phys.}, 15(2):025033, 2013.

\bibitem{Kurz2015}
C.~Kurz.
\newblock {\em Quantum networking with single ions and single photons
  interfaced in free space}.
\newblock PhD thesis, Universit\"at des Saarlandes, Saarbr\"ucken, 2015.

\bibitem{Barros2009}
H.~G. Barros, A.~Stute, T.~E. Northup, C.~Russo, P.~O. Schmidt, and R.~Blatt.
\newblock Deterministic single-photon source from a single ion.
\newblock {\em New J. Phys.}, 11(10):103004, 2009.

\bibitem{Hadfield2009}
R.~H. Hadfield.
\newblock Single-photon detectors for optical quantum information applications.
\newblock {\em Nature Photon}, 3(12):696--705, Dec 2009.

\bibitem{Dusek2000}
M.~Dusek, M.~Jahma, and N.~L\"utkenhaus.
\newblock Unambiguous state discrimination in quantum cryptography with weak
  coherent states.
\newblock {\em Phys. Rev. A}, 62:022306, Jul 2000.

\bibitem{Sangouard2011}
N.~Sangouard, C.~Simon, H.~de~Riedmatten, and N.~Gisin.
\newblock Quantum repeaters based on atomic ensembles and linear optics.
\newblock {\em Rev. Mod. Phys.}, 83:33--80, Mar 2011.

\bibitem{Ma1994}
L.-S. Ma, P.~Jungner, J.~Ye, and J.~L. Hall.
\newblock Delivering the same optical frequency at two places: accurate
  cancellation of phase noise introduced by an optical fiber or other
  time-varying path.
\newblock {\em Opt. Lett.}, 19(21):1777, 1994.

\bibitem{Minar2008}
J.~Min{\`a}\v{r}, H.~de~Riedmatten, C.~Simon, H.~Zbinden, and N.~Gisin.
\newblock Phase-noise measurements in long-fiber interferometers for
  quantum-repeater applications.
\newblock {\em Phys. Rev. A}, 77(5), May 2008.

\bibitem{Jiang2008}
H.~Jiang, F.~Kéfélian, S.~Crane, O.~Lopez, M.~Lours, J.~Millo, D.~Holleville,
  P.~Lemonde, C.~Chardonnet, A.~Amy-Klein, and et~al.
\newblock Long-distance frequency transfer over an urban fiber link using
  optical phase stabilization.
\newblock {\em J. Opt. Soc. Am. B}, 25(12):2029, 2008.

\bibitem{Duan2000}
L.-M. Duan, G.~Giedke, J.~I. Cirac, and P.~Zoller.
\newblock Entanglement purification of gaussian continuous variable quantum
  states.
\newblock {\em Phys. Rev. Lett.}, 84:4002--4005, Apr 2000.

\bibitem{Pan2001}
J.-W. Pan, C.~Simon, C.~Brukner, and A.~Zeilinger.
\newblock Entanglement purification for quantum communication.
\newblock {\em Nature}, 410(6832):1067--1070, April 2001.

\bibitem{Haeffner2005}
H.~H{\"a}ffner, F.~Schmidt-Kaler, W.~H{\"a}nsel, C.~Roos, T.~K{\"o}rber,
  M.~Chwalla, M.~Riebe, J.~Benhelm, U.~Rapol, C.~Becher, and R.~Blatt.
\newblock Robust entanglement.
\newblock {\em Appl. Phys. B}, 81(2-3):151--153, 2005.

\bibitem{Schindler2011}
P.~Schindler, J.~T. Barreiro, T.~Monz, V.~Nebendahl, D.~Nigg, M.~Chwalla,
  M.~Hennrich, and R.~Blatt.
\newblock Experimental repetitive quantum error correction.
\newblock {\em Science}, 332(6033):1059--1061, May 2011.

\end{thebibliography}
  \bibliographystyle{unsrt}

%
%

\end{document}